\documentclass[floatfix,raggedbottom,reprint,amsmath,amssymb,aip,jap,showpacs]{revtex4-1}
\usepackage{epsfig}
\usepackage{dcolumn}
\usepackage{bm}
\usepackage{fancyhdr}
\usepackage{textcomp}
\usepackage{hhline}
\usepackage{calc}
\usepackage[labelformat=simple]{subcaption}
\usepackage{amsfonts}
\usepackage{booktabs}
\graphicspath{{.}}

\newcolumntype{L}[1]{>{\raggedright\arraybackslash}p{#1}}
\newcolumntype{C}[1]{>{\centering\arraybackslash}p{#1}}
\newcolumntype{R}[1]{>{\raggedleft\arraybackslash}p{#1}}
\newlength{\myheight}
\begin{document}
\title{Topologically nontrivial phases in superconducting transition metal carbides}
\author{Richard Zhan}
\author{Xuan Luo}
\affiliation{National Graphene Research and Development Center, Springfield, Virginia 22151, USA}
\date{\today}
\begin{abstract}
Topological superconductors have shown great potential in the search for unique quasiparticles such as Majorana fermions.
Combining nontrivial band topology and superconductivity can lead to topological superconductivity due to the proximity effect.
In this work, we used first principle calculations to predict that rock-salt phases of VC and CrC are superconducting with topologically nontrivial states.
The phonon dispersions of these transition metal carbides displayed no imaginary frequencies, which suggests dynamic stability.
Additionally, the presence of soft acoustic phonon bands supports the existence of Bardeen-Cooper-Schrieffer (BCS) superconductivity in rock-salt VC and CrC.
Therefore, these transition metal carbides are practical candidates for studying topological superconductors and their associated Majorana bound states.

\end{abstract}
\newpage
\maketitle

\newlength{\mywidth}
\setlength{\mywidth}{\linewidth-\columnsep}

\section{Introduction}

Quantum computing has gained interest in recent years due to its potential to disrupt many different fields including biology \cite{perdomo2012finding}, chemistry \cite{lanyon2010towards}, and computer science \cite{li2001quantum}.
One of the biggest challenges of creating a practical quantum computer is the effect of environmental noise on its calculations.
Topological quantum computers are a possible solution to this issue.
These devices use Majorana fermions (MFs) to encode information that is invariant to local perturbations \cite{nayak2008non-abelian}.
However, realizing MFs has proven difficult since the initial prediction of their existence in 1937 by Ettore Majorana.
On the other hand, MFs have been predicted to bind to defects in certain superconductors (SC), creating Majorana bound states (MBS) \cite{wilczek2009majorana}.

One way to observe these MBSs is with topological superconductors (TSC).
MBS exist in the normal core of quantum vortices as well as on the boundaries of magnetic domains in TSCs.
Therefore, studying MBSs in TSCs may help advance the creation of practical quantum computers \cite{zhang2018observation}.
Previous studies have identified multiple ways to achieve topological superconductivity.
One approach to topological superconductivity uses proximity effect induced superconductors.
When a topological insulator or semiconductor with Rashba spin-orbit coupling (SOC) is brought into contact with an s-wave SC, the interface turns into a TSC \cite{fu2008superconducting, lutchyn2010majorana}.
Previous studies have produced promising results for these topologically superconducting heterostructures (TSH).
In 2014, the first TSH was experimentally realized \cite{xu2014artificial}.
One year later, researchers observed MBSs in the same TSH of Bi$_2$Te$_3$/NbSe$_2$ \cite{xu2015experimental}.
TSHs, however, can be complex and consequently difficult to create, which hinders further research into these types of TSCs.
In addition, most TSHs have relatively low SC transition temperatures, usually below 4 K \cite{huang2018multiple}.

Recent work has successfully brought together the two fundamental aspects of TSCs in a single material, the topological nature of the band structure and superconductivity.
Materials such as Au$_2$Pb \cite{schoop2015dirac, xing2016superconductivity}, CaSn$_3$ \cite{gupta2017topologically}, Mg$_2$Pb \cite{bian2017prediction}, and MoC \cite{huang2018multiple} have been predicted to be superconducting and have topologically nontrivial band structures.
$\alpha$-MoC, a transition metal carbide, was one of the most promising materials due to its relatively high T$_{\text{c}}$ of 14 K.

In this study, we use first-principle calculations to determine the electronic band structure of the superconductors VC, CrC, NbC, and TaC.
We show that VC and CrC are potential candidates for studying topologically superconductivity by calculating their $\mathbb{Z}_2$ topological invariants.
In addition, the phonon band structures were calculated to show the stability and superconductivity of these materials.
Afterwards, we discuss methods of synthesizing VC and CrC.
Thus, both of these compounds are promising candidates for studying topological superconductors and their associated MBSs.

Our methods for our first principle calculations can be found in Section II.
We present and discuss our results on the topological superconductivity of VC and CrC in Section III.
Finally, our conclusion is found in Section IV.

\section{Methodology}

\subsection{Computational Details}

The density functional theory \cite{hohenberg1964inhomogeneous, kohn1965self-consistent} (DFT) calculations were performed with ABINIT \cite{gonze2016recent, gonze2009abinit, gonze2005a}.
We used generalized gradient approximation (GGA) Perdew-Burke-Ernzerhof (PBE) functionals \cite{perdew1996generalized} and projector augmented wave (PAW) pseudopotentials \cite{torrent2008implementation, blochl1994projector, kresse1999from} in our calculations.
The PAW pseudopotentials were generated using AtomPAW \cite{holzwarth2001a}.
The electron configuration and cutoff radius for the pseudopotentials of each element used in our calculations are shown in Table \ref{elements}.
We used the Broyden-Fletcher-Goldfarb-Shanno (BFGS) minimization \cite{broyden1970the, fletcher1970a, goldfarb1970a, shanno1970conditioning, shanno1970optimal} for structural optimization.
The $\mathbb{Z}_2$ topological invariants and Wilson loops were calculated with ABINIT and Z2Pack \cite{gresch2017z2pack, soluyanov2011wannier}.
The phonon frequencies were calculated using density functional perturbation theory (DFPT) \cite{baroni2001phonons}.

In strongly correlated systems such as superconductors not explained by the BCS theory of superconductivity, either DFT+U or dynamical mean field theory (DMFT) is used, since GGA and local density approximation (LDA) are generally not accurate enough \cite{liechtenstein1995density, kotliar2006electronic}. 
However, the superconductivity of VC, CrC, NbC, and TaC follows the BCS theory, so these materials are not strongly correlated systems. 
Additionally, many previous studies have succesfully used GGA and LDA exchange correlation functionals to compute the properties of similar superconducting transition metal carbides \cite{huang2018multiple, kavitha2016structural, tutuncu2012electrons, antonin2001correlation, isaev2007phonon}. 
Consequently, we do not need to make use of DFT+U or DMFT for our calculations.

\begin{table}[ht]
	\centering
	\caption{Each element's valence electron configuration and radius cutoff used in generating the PAW pseudopotentials.}
	\begin{tabular}{L{0.15\mywidth}C{0.425\mywidth}C{0.425\mywidth}}
		\hhline{===}
		Element & Valence Configuration & Cutoff Radius (Bohr) \\
		\hline
		C & $2s^22p^2$ & 1.51 \\
		V & $3s^24s^23p^63d^3$ & 2.20 \\
		Cr & $3s^24s^13p^63d^5$ & 2.11 \\
		Nb & $4s^25s^14p^64d^4$ & 2.21 \\
		Ta & $5s^26s^25p^65d^3$ & 2.41 \\
		\hhline{===}
	\end{tabular}
	\label{elements}
\end{table}

\subsection{Total Energy Convergence}

The plane-wave kinetic energy cutoff and k-point grid of each compound were converged after they reached the tolerance criteria of $10^{-4}$ Ha twice successively.
The energy cutoff and k-mesh we used for each compound can be found in Table \ref{convergence}.
We used a broadening factor ($\sigma$) of 0.005 Ha for the Gaussian smearing described by the following equation.

\begin{equation}
        \begin{array}{c}
                \widetilde{\delta}(x) = \frac{e^{-x^2}}{sqrt(\pi)}, \ x = \frac{\epsilon-\epsilon_F}{\sigma}
        \end{array}
\end{equation}

\begin{table}[ht]
	\centering
	\caption{Each compound's converged kinetic energy cutoff and k-point grid for the GGA and LDA pseudopotentials}
	\begin{ruledtabular}
		\begin{tabular}{L{0.25\mywidth}C{0.375\mywidth}C{0.375\mywidth}}
			Compound & Energy Cutoff (Ha) & K-Point Grid \\
			\hline
			VC & 17\footnotemark[1] & $6 \times 6 \times 6$ \\
			CrC & 23 & $6 \times 6 \times 6$ \\
			NbC & 24 & $14 \times 14 \times 14$ \\
			TaC & 24 & $18 \times 18 \times 18$ \\
		\end{tabular}
	\end{ruledtabular}
	\footnotetext[1]{All of the convergence tests resulted in the same values for both GGA and LDA pseudopotentials except for kinetic energy cutoff of VC. It was determined to be 17 Ha for GGA and 16 Ha for LDA.
}
	\label{convergence}
\end{table}

\subsection{Electronic Band Structure Methodology}

\begin{figure}[ht]
	\begin{subfigure}[t]{0.5\mywidth}
		\caption{}
		\label{unitcell}
		\includegraphics[height=3.25cm, clip=true]{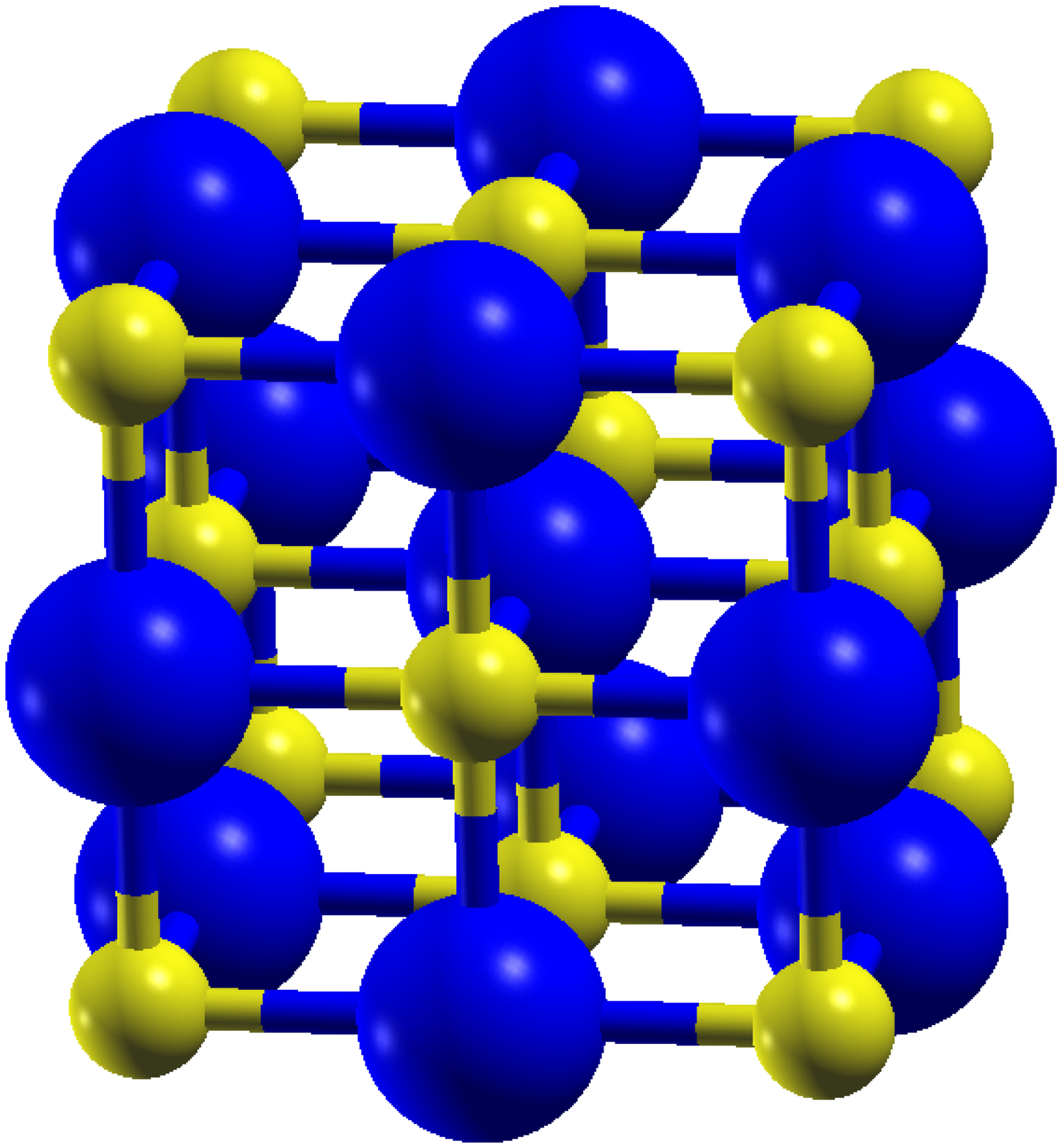}
	\end{subfigure}
	\begin{subfigure}[t]{0.5\mywidth}
		\caption{}
		\label{brillouin}
		\includegraphics[height=3.25cm, clip=true]{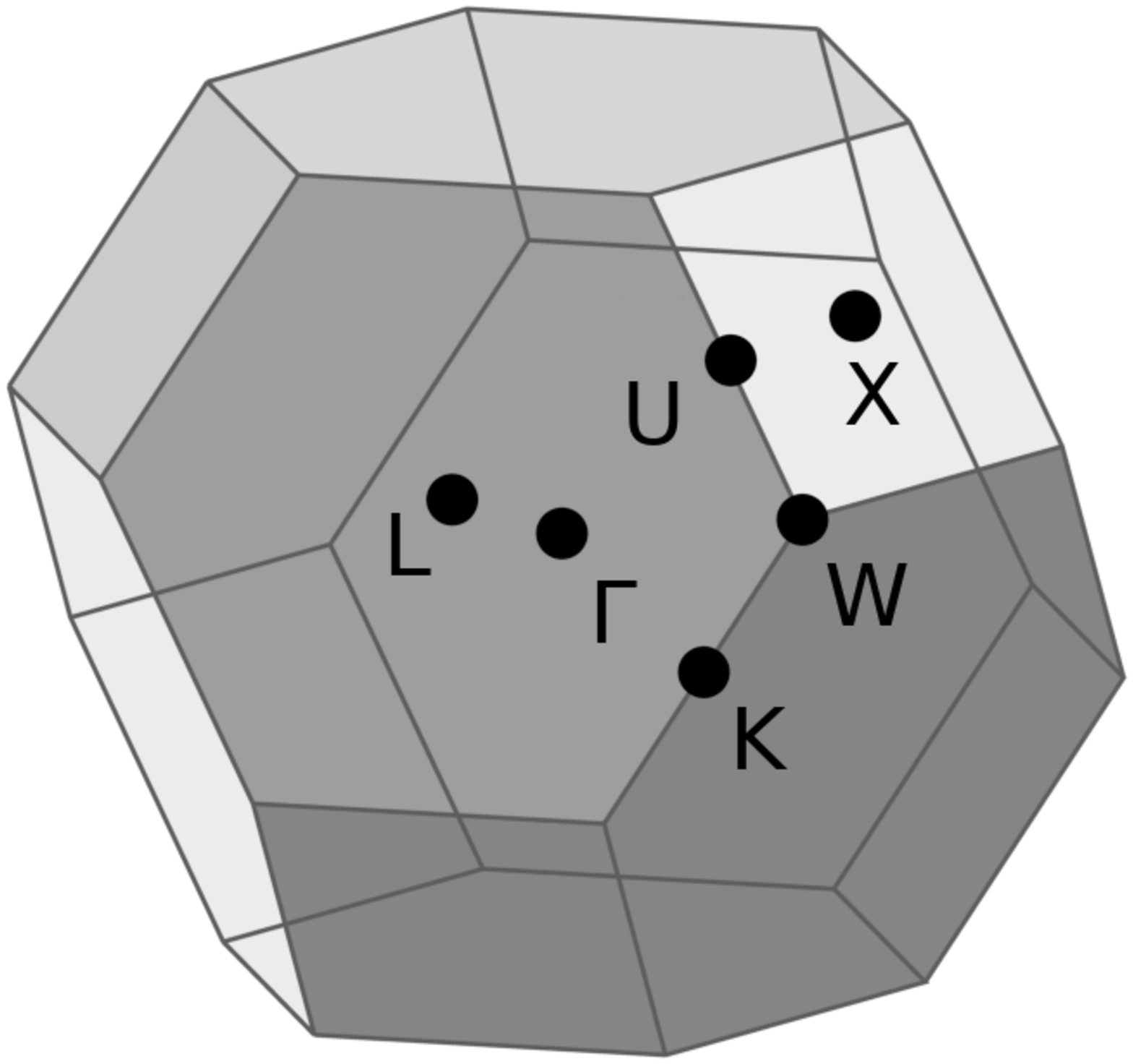}
	\end{subfigure}
	\caption{
	(a) The rock-salt structure unit cell. 
	The transition metal atoms (V, Cr, Nb, Ta) are represented by the large blue atoms. 
	The carbon atoms are represented by the small yellow atoms.
	(b) The first Brillouin zone and high symmetry k-points of the rock-salt structure.
	}
\end{figure}

We used the high-symmetry points $\Gamma$ (0, 0, 0), X (0, 0.5, 0.5), L (0.5, 0.5, 0.5), W (0.25, 0.75, 0.5), U (0.25, 0.625, 0.625), and K (0.375, 0.75, 0.375) shown in Figure \ref{brillouin}.
Our k-point sampling \cite{monkhorst1976special} was generated with the shifts (0.5, 0.5, 0.5), (0.5, 0, 0), (0, 0.5, 0), and (0, 0, 0.5).
We calculated band structures with and without SOC for each material.

\subsection{Phonon Dispersion Methodology}

DFPT is mainly used to compute the derivative of the electronic energy of a system with respect to certain perturbations.
Vibrational, dielectric, and piezoelectric properties can be calculated through the use of DFPT \cite{baroni2001phonons}.
We used ABINIT's implementation of DFPT to perform lattice-dynamical calculations with the linear response method \cite{gonze2005first}.
This linear response approach allows us to calculate the phonon dispersions through the use of a single unit cell, eliminating the need of a supercell construction.
For our calculations, we used a $16 \times 16 \times 16$ k-point grid and an $8 \times 8 \times 8$ q-point grid.
All of our DFPT calculations used fully relaxed structures and LDA plane-wave exchange correlation functionals.

\begin{figure*}[ht]
	\centering
	\begin{subfigure}[b]{0.245\textwidth}
		\caption{}
		\label{vcall}
		\includegraphics[height=\myheight, clip=true]{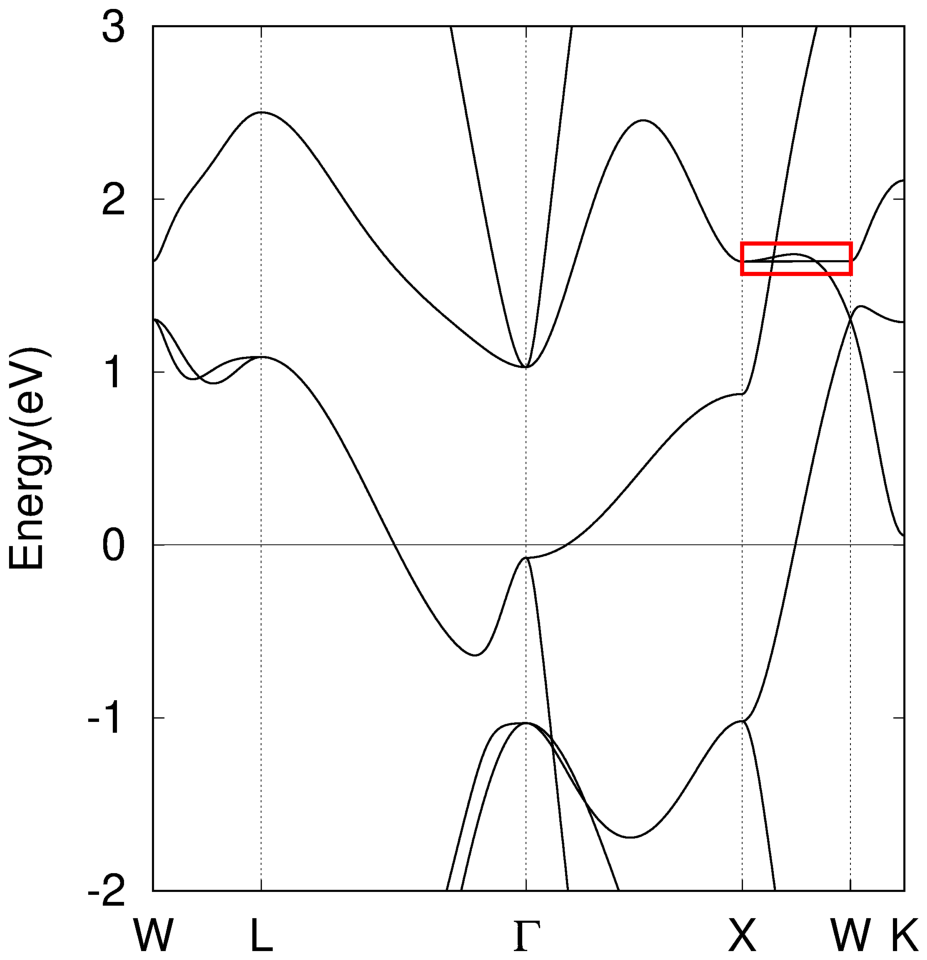}
	\end{subfigure}
	\hfill
	\begin{subfigure}[b]{0.33\textwidth}
		\caption{}
		\label{vcsocall}
		\includegraphics[height=\myheight, clip=true]{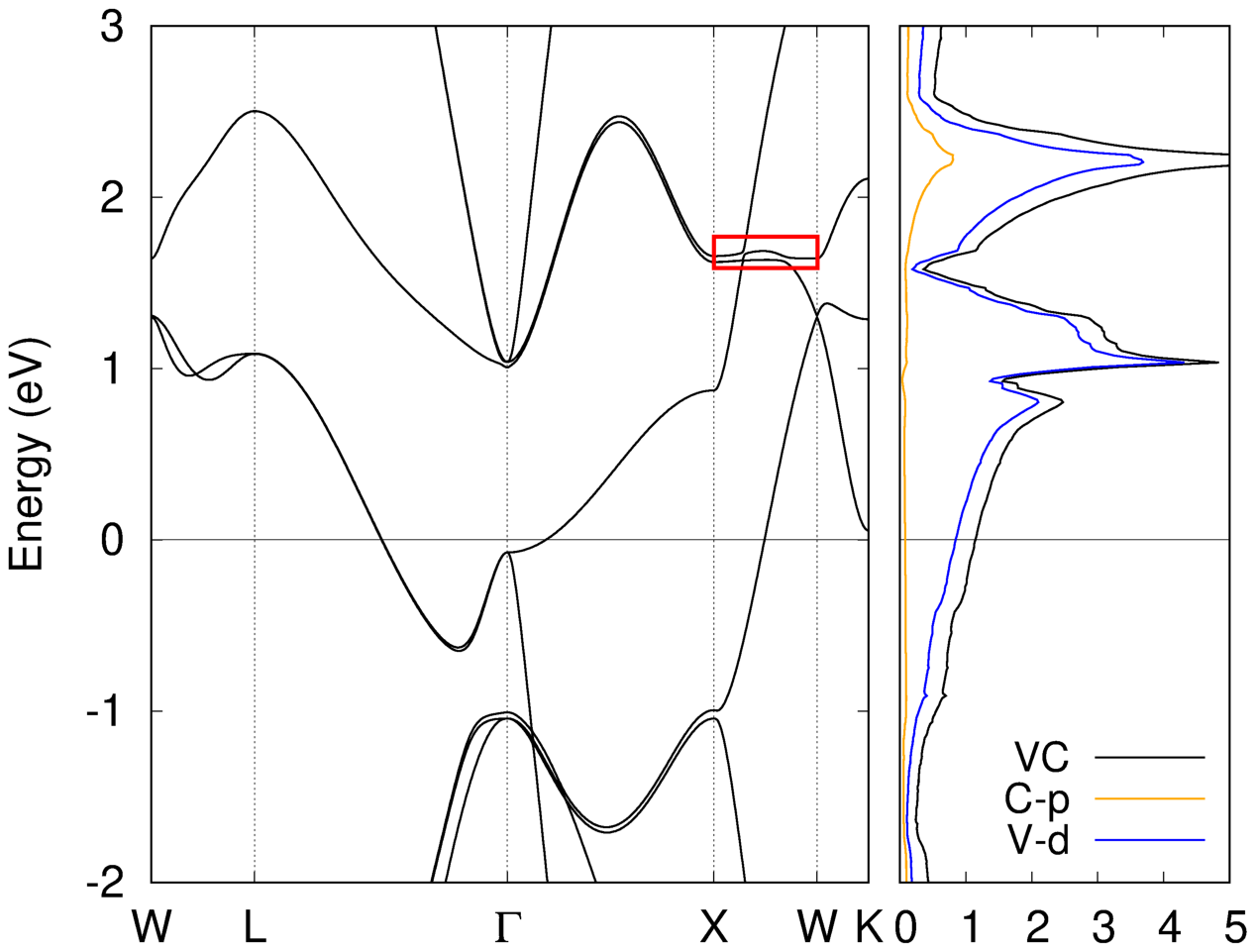}
	\end{subfigure}
	\hfill
	\begin{subfigure}[b]{0.2\textwidth}
		\caption{}
		\label{vcxw}
		\includegraphics[height=\myheight, clip=true]{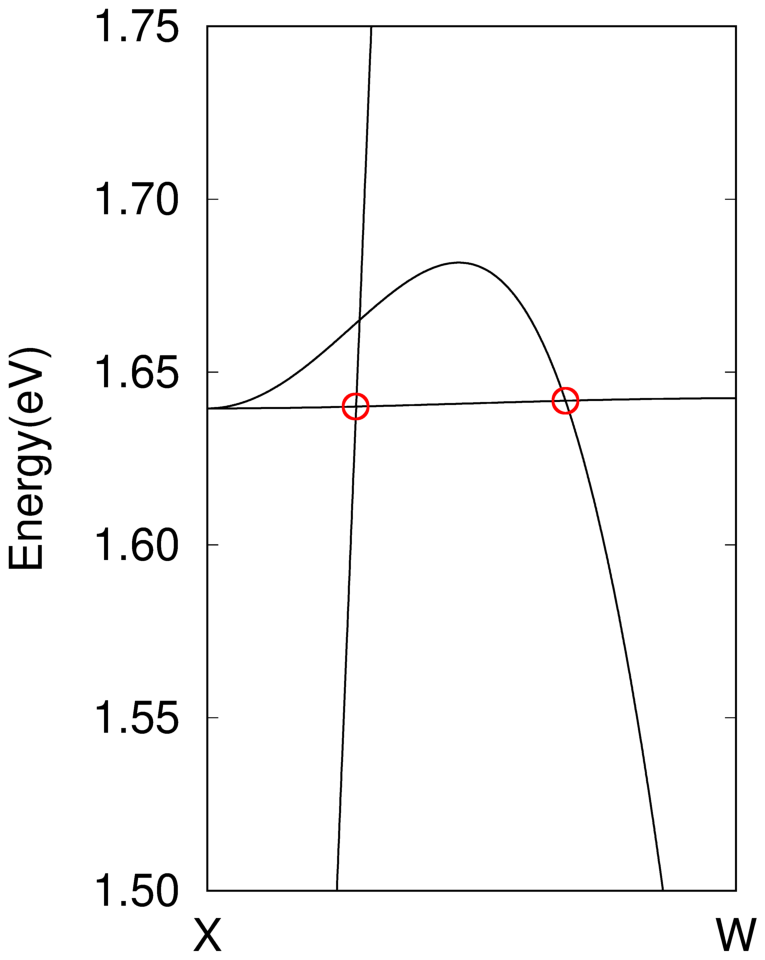}
	\end{subfigure}
	\hfill
	\begin{subfigure}[b]{0.2\textwidth}
		\caption{}
		\label{vcsocxw}
		\includegraphics[height=\myheight, clip=true]{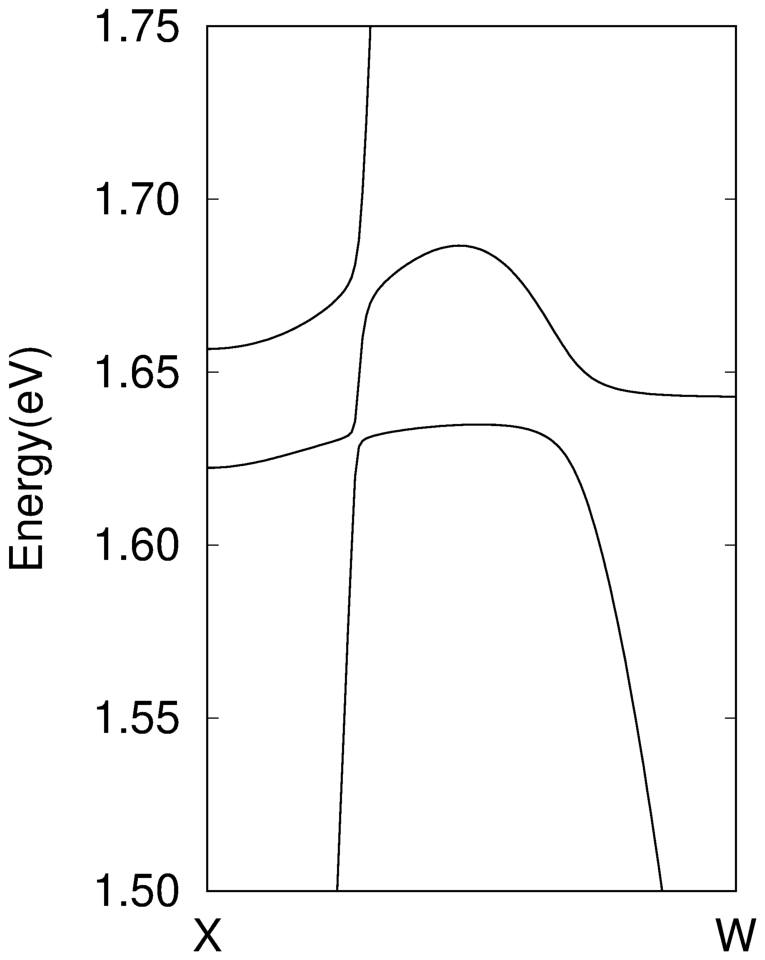}
	\end{subfigure}
	\\
	\begin{subfigure}[b]{0.245\textwidth}
		\caption{}
		\label{crcall}
		\includegraphics[height=\myheight, clip=true]{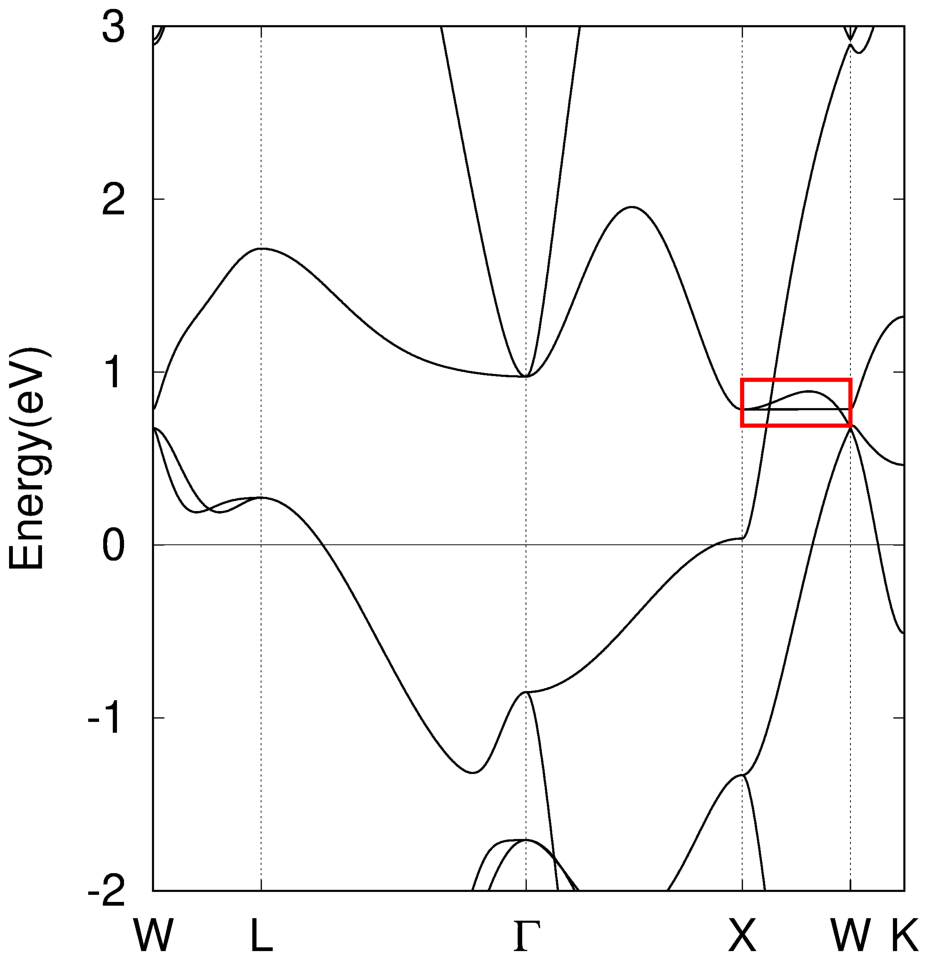}
	\end{subfigure}
	\hfill
	\begin{subfigure}[b]{0.33\textwidth}
		\caption{}
		\label{crcsocall}
		\includegraphics[height=\myheight, clip=true]{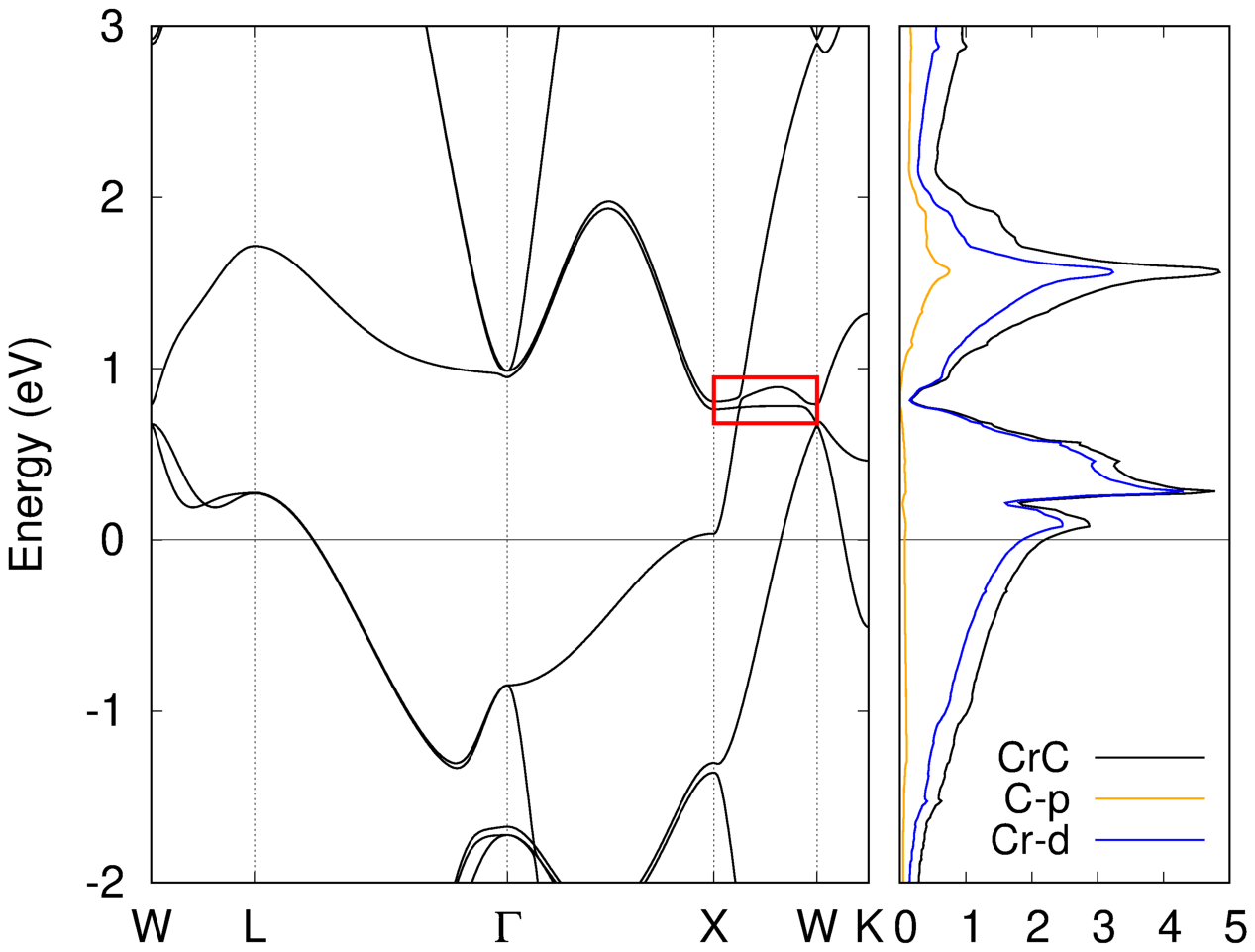}
	\end{subfigure}
	\hfill
	\begin{subfigure}[b]{0.2\textwidth}
		\caption{}
		\label{crcxw}
		\includegraphics[height=\myheight, clip=true]{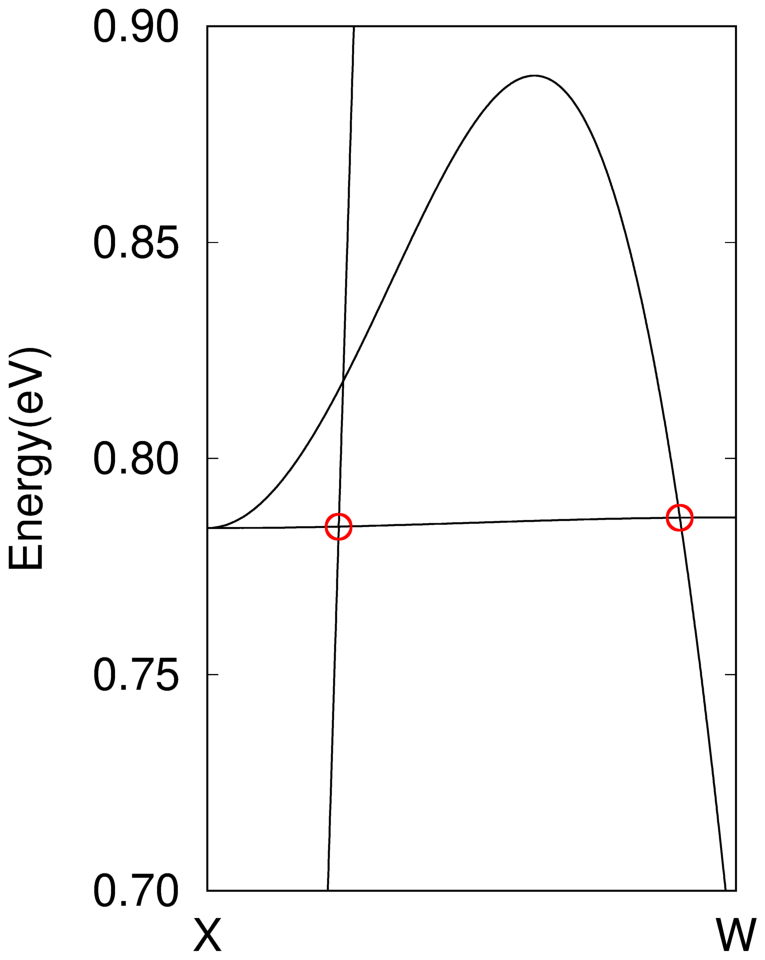}
	\end{subfigure}
	\hfill
	\begin{subfigure}[b]{0.2\textwidth}
		\caption{}
		\label{crcsocxw}
		\includegraphics[height=\myheight, clip=true]{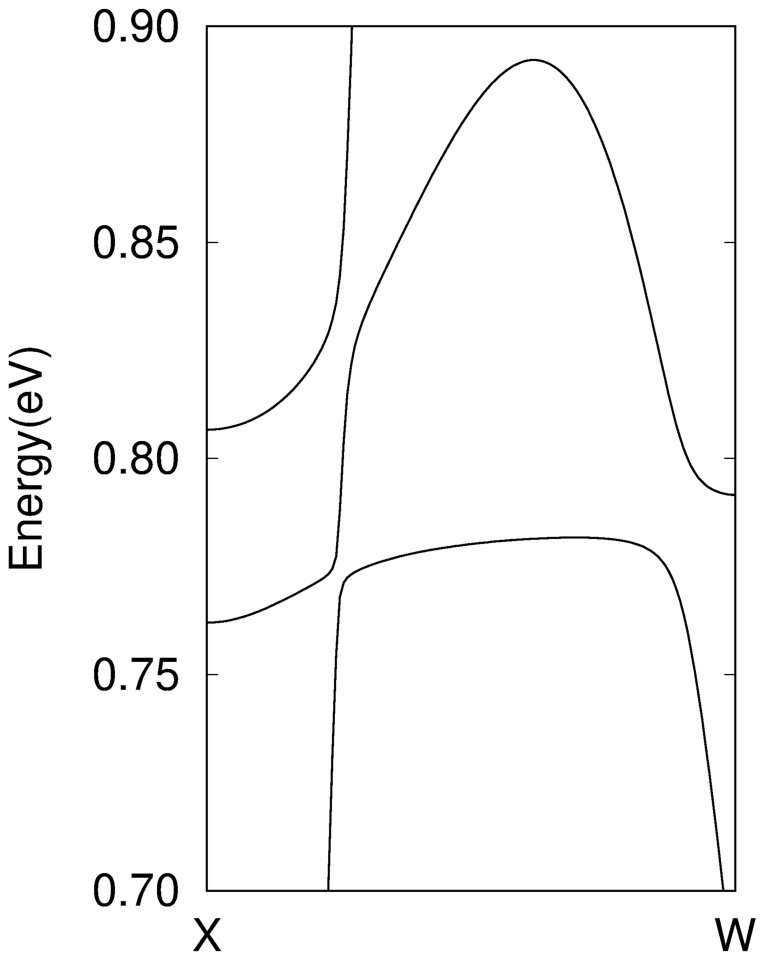}
	\end{subfigure}
	\caption{
	(a) The band structure of VC without SOC.
	(b) The band structure and DOS of VC with SOC.
	(c) Magnified view of the red box in (a). The two red circles denote the nodal ring.
	(d) Magnified view of the red box in (b).
	(e) The band structure of CrC without SOC.
	(f) The band structure and DOS of CrC with SOC.
	(g) Magnified view of the red box in (e). The two red circles denote the nodal ring.
	(h) Magnified view of the red box in (f).}
\end{figure*}

\begin{figure*}
	\begin{subfigure}[b]{0.245\textwidth}
		\caption{}
		\label{nbcall}
		\includegraphics[height=\myheight, clip=true]{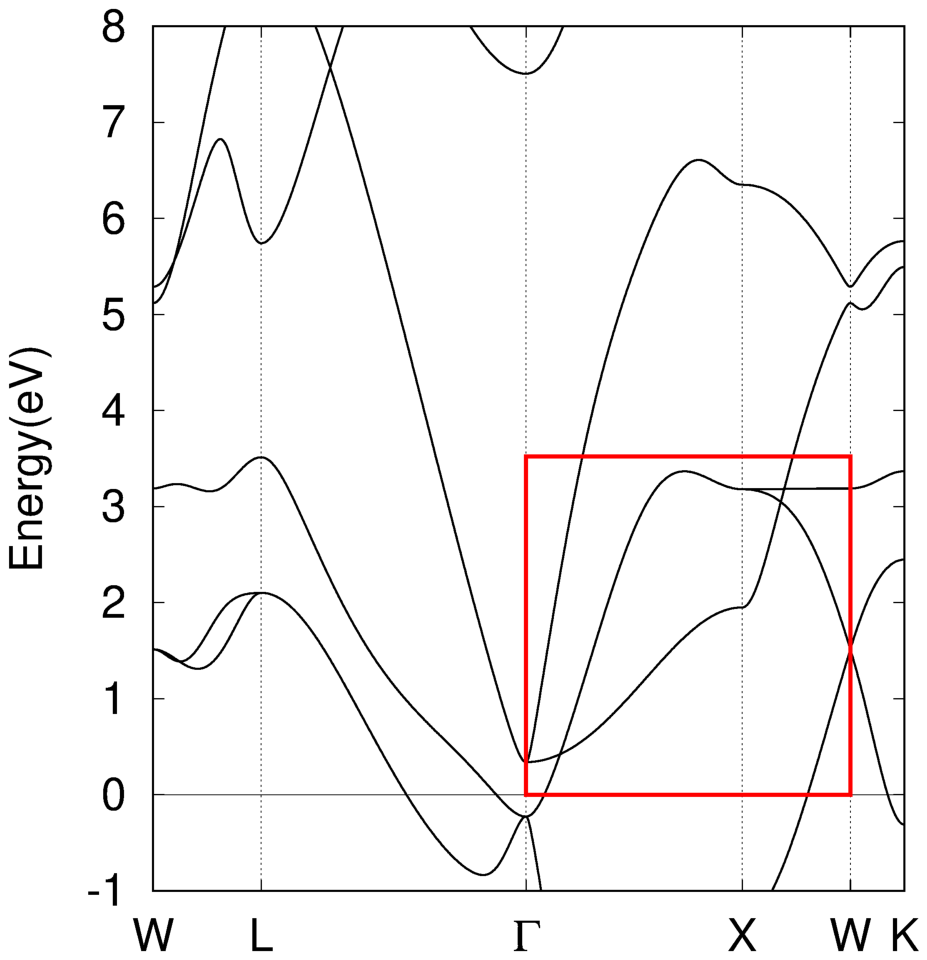}
	\end{subfigure}
	\hfill
	\begin{subfigure}[b]{0.33\textwidth}
		\caption{}
		\label{nbcsocall}
		\includegraphics[height=\myheight, clip=true]{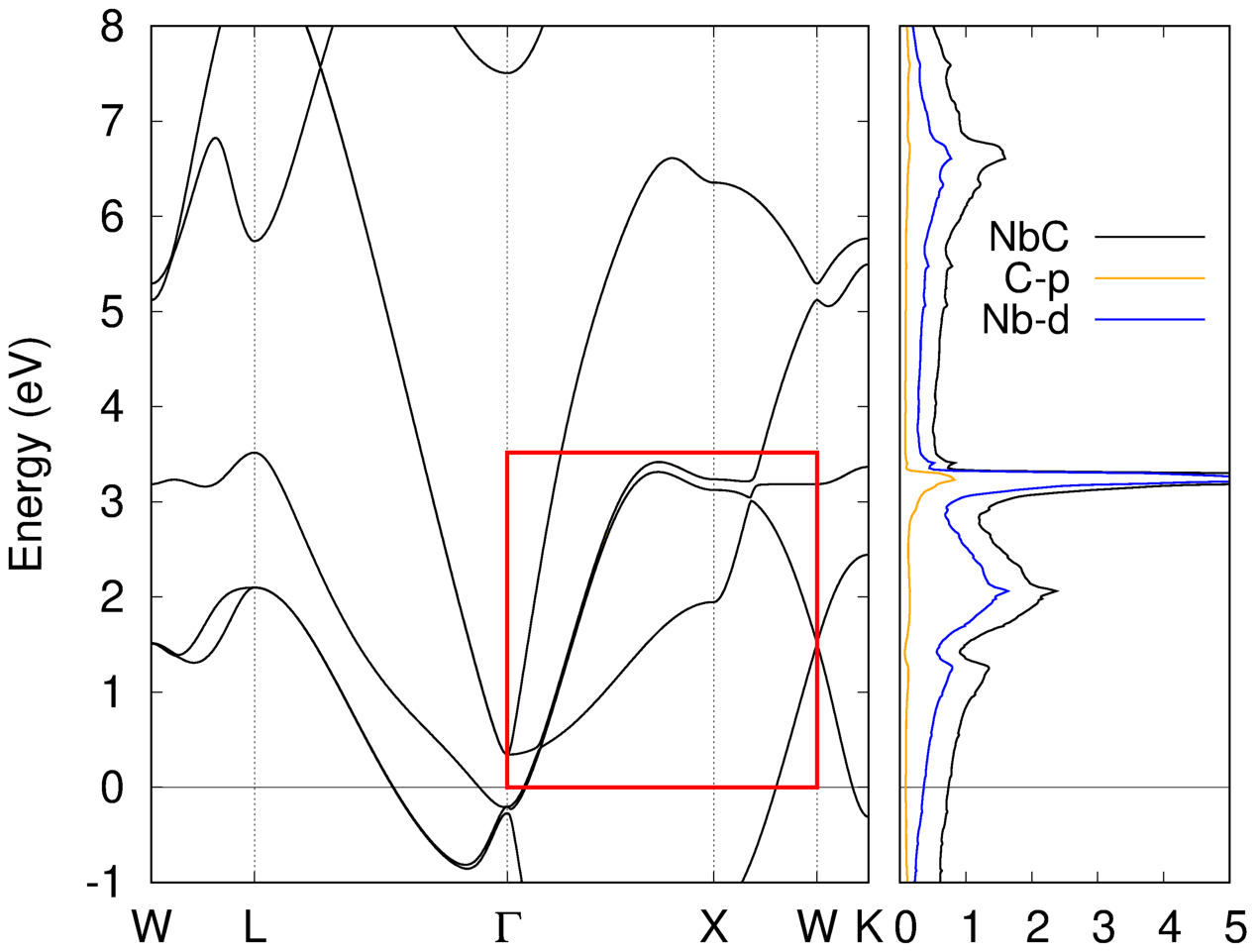}
	\end{subfigure}
	\hfill
	\begin{subfigure}[b]{0.2\textwidth}
		\caption{}
		\label{nbcgxw}
		\includegraphics[height=\myheight, clip=true]{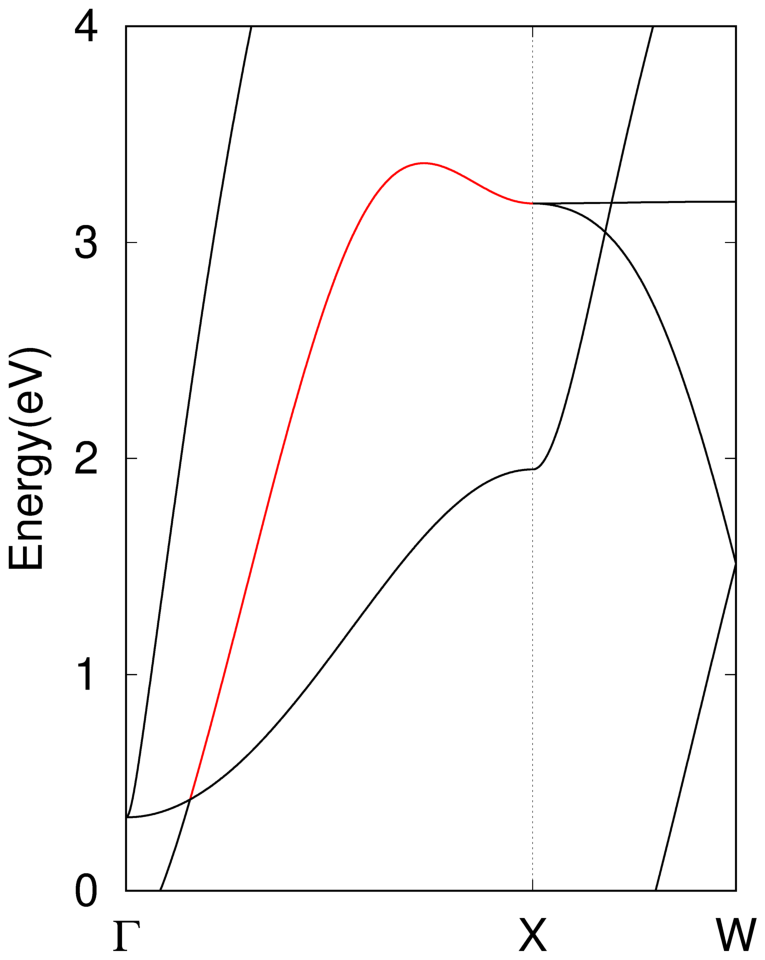}
	\end{subfigure}
	\hfill
	\begin{subfigure}[b]{0.2\textwidth}
		\caption{}
		\label{nbcsocgxw}
		\includegraphics[height=\myheight, clip=true]{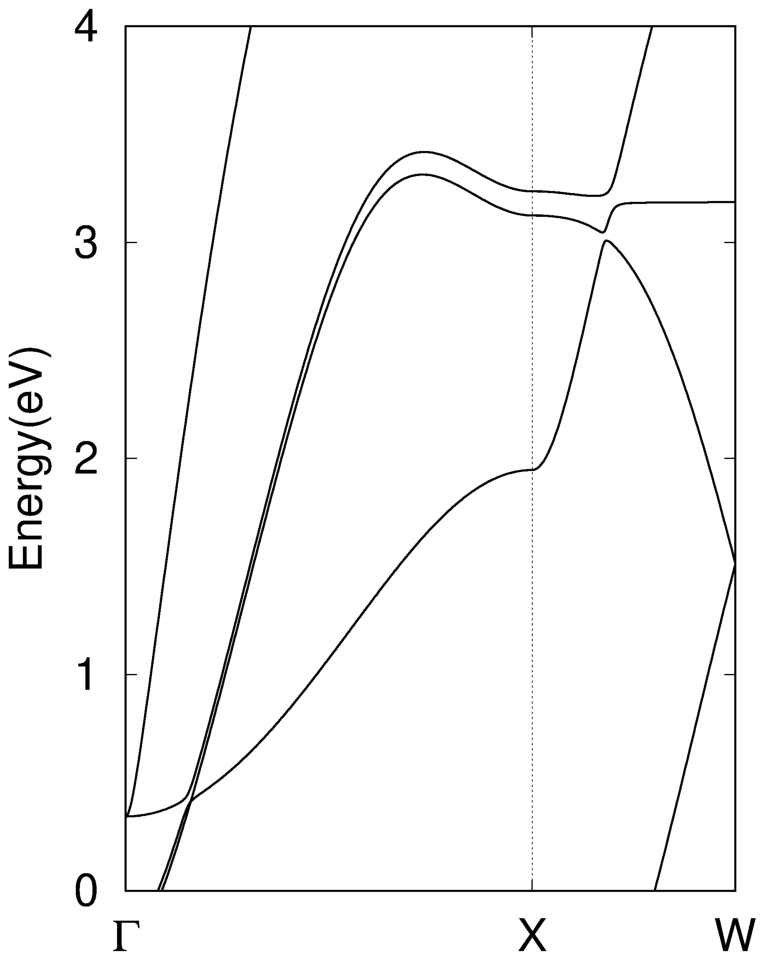}
	\end{subfigure}
	\\
	\begin{subfigure}[b]{0.245\textwidth}
		\caption{}
		\label{tacall}
		\includegraphics[height=\myheight, clip=true]{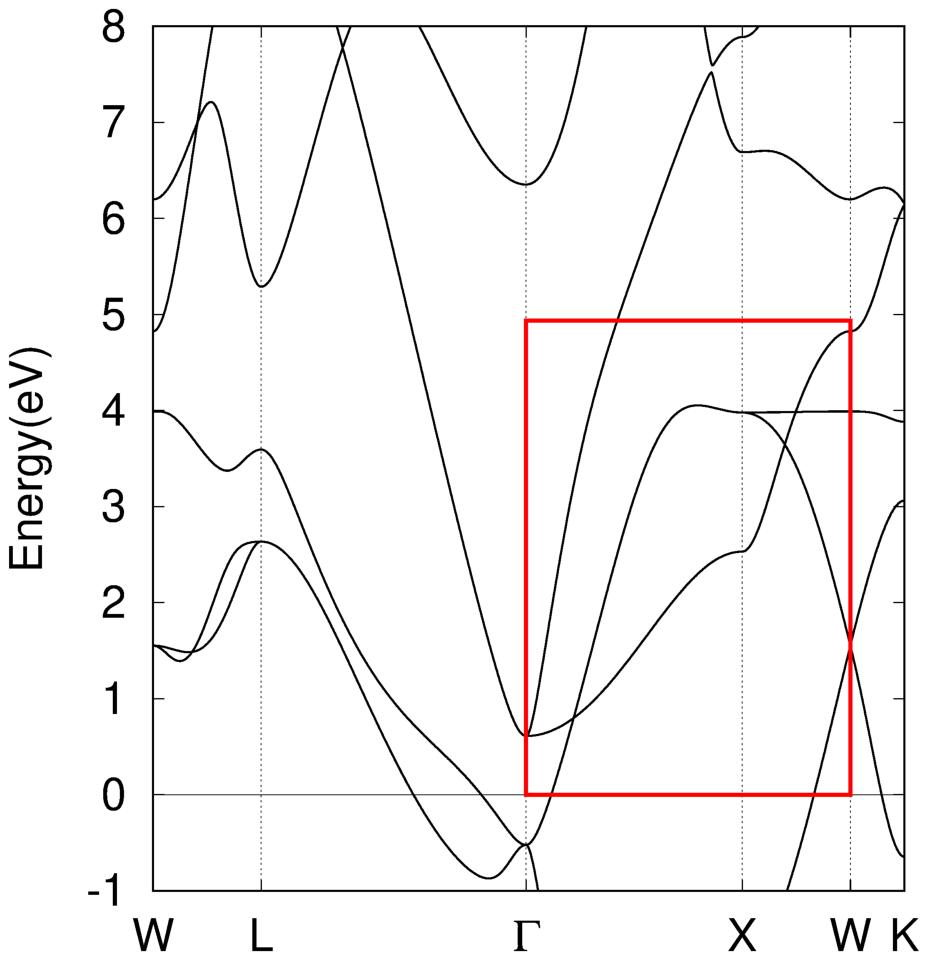}
	\end{subfigure}
	\hfill
	\begin{subfigure}[b]{0.33\textwidth}
		\caption{}
		\label{tacsocall}
		\includegraphics[height=\myheight, clip=true]{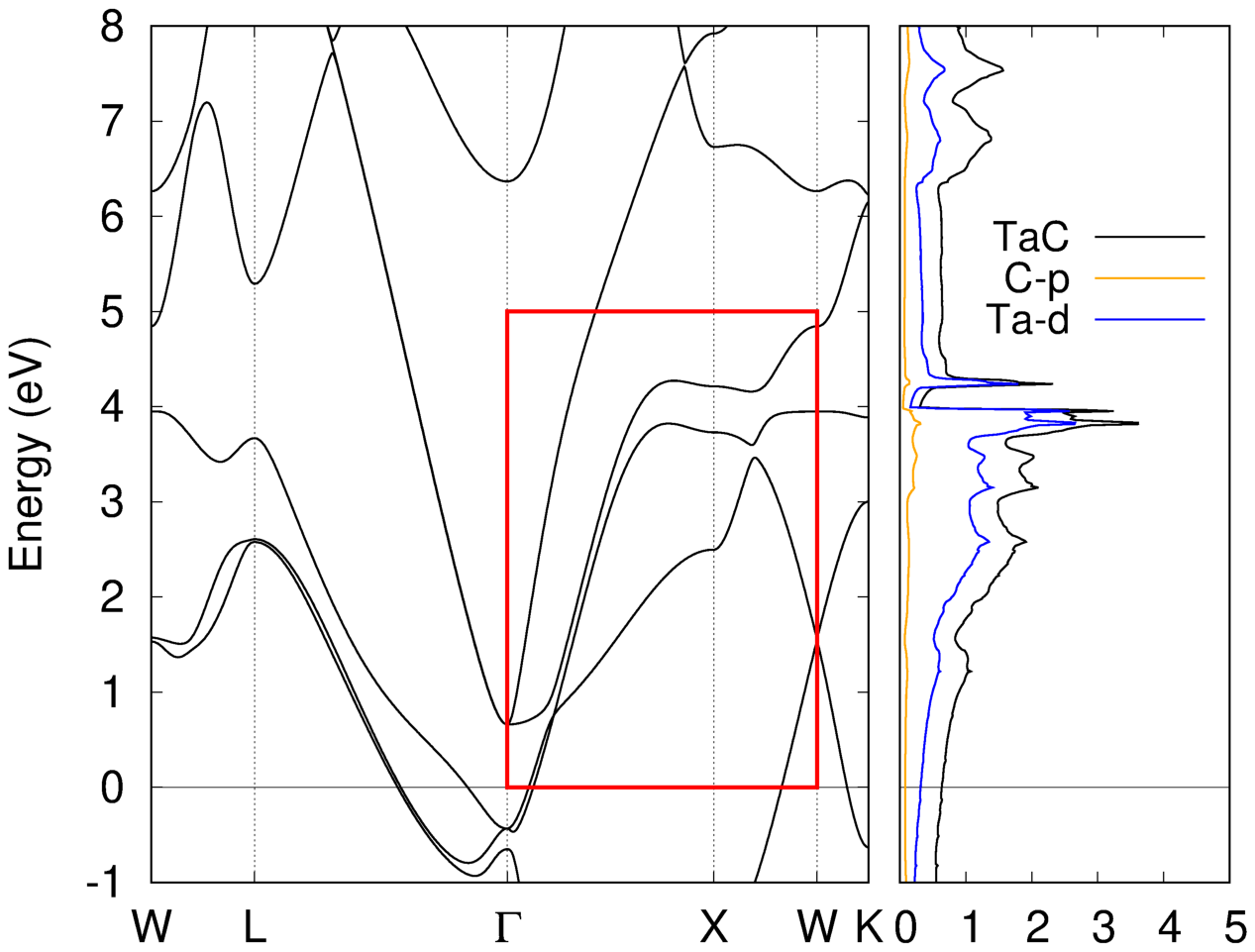}
	\end{subfigure}
	\hfill
	\begin{subfigure}[b]{0.2\textwidth}
		\caption{}
		\label{tacgxw}
		\includegraphics[height=\myheight, clip=true]{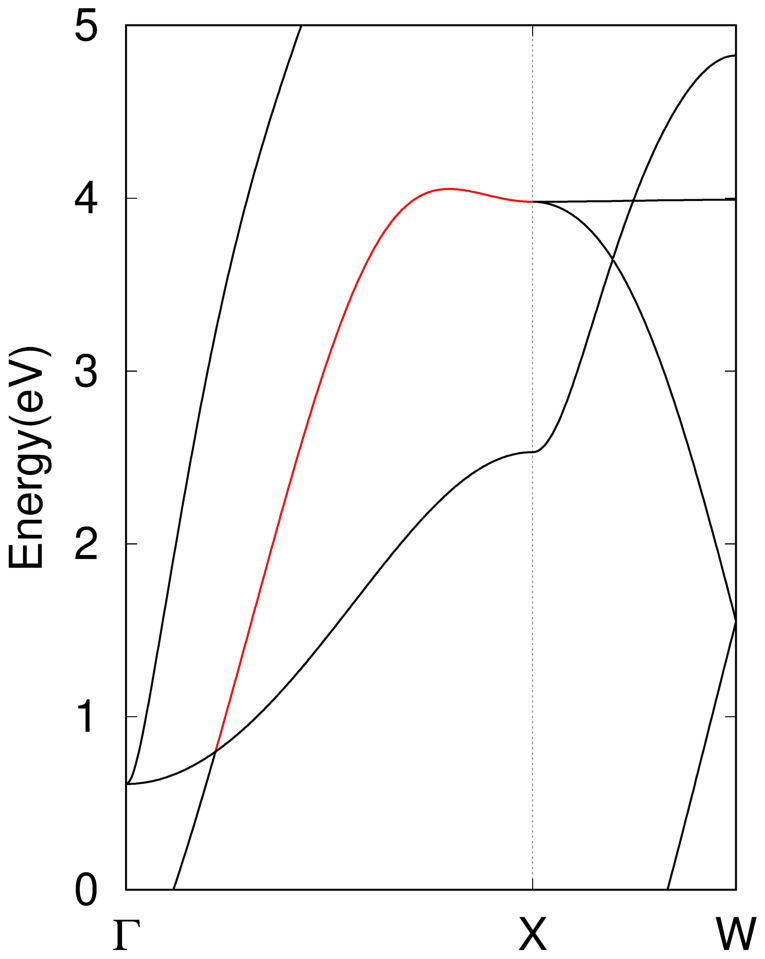}
	\end{subfigure}
	\hfill
	\begin{subfigure}[b]{0.2\textwidth}
		\caption{}
		\label{tacsocgxw}
		\includegraphics[height=\myheight, clip=true]{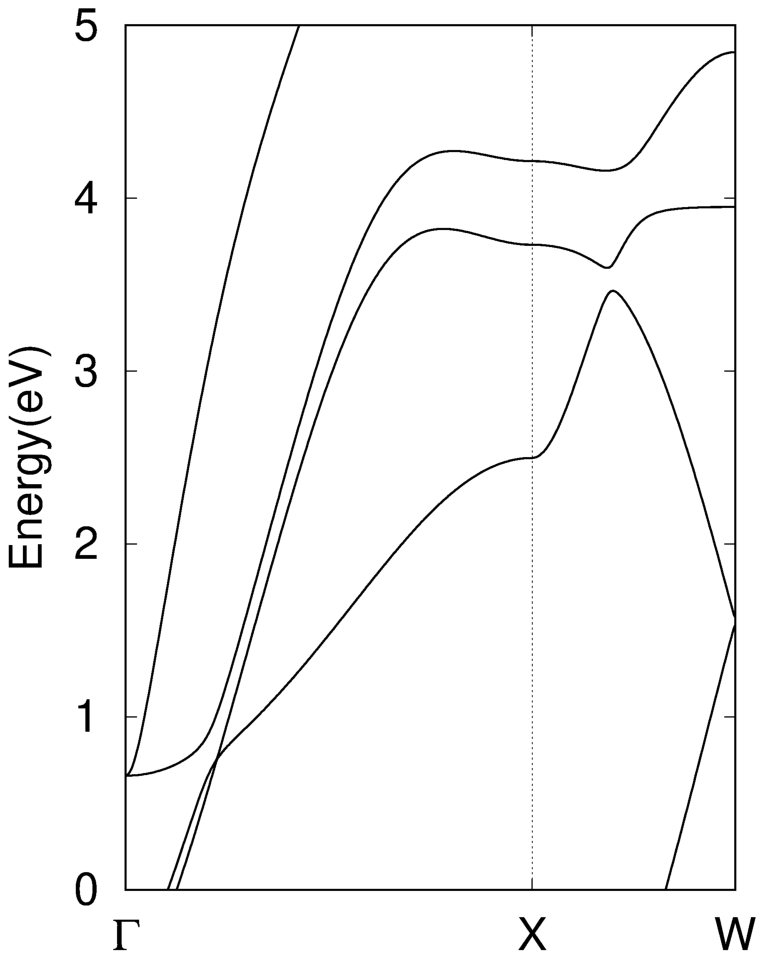}
	\end{subfigure}
	\caption{
	(a) The band structure of NbC without SOC.
	(b) The band structure and DOS of NbC with SOC.
	(c) Magnified view of the red box in (a). The red line denotes the doubly degenerate bands.
	(d) Magnified view of the red box in (b).
	(e) The band structure of TaC without SOC.
	(f) The band structure and DOS of TaC with SOC.
	(g) Magnified view of the red box in (e). The red line denotes the doubly degenerate bands.
	(h) Magnified view of the red box in (f).}
\end{figure*}

\begin{figure*}[ht]
        \centering
        \begin{subfigure}[b]{0.24\textwidth}
                \caption{}
                \label{ycz2a}
                \includegraphics[width=\textwidth]{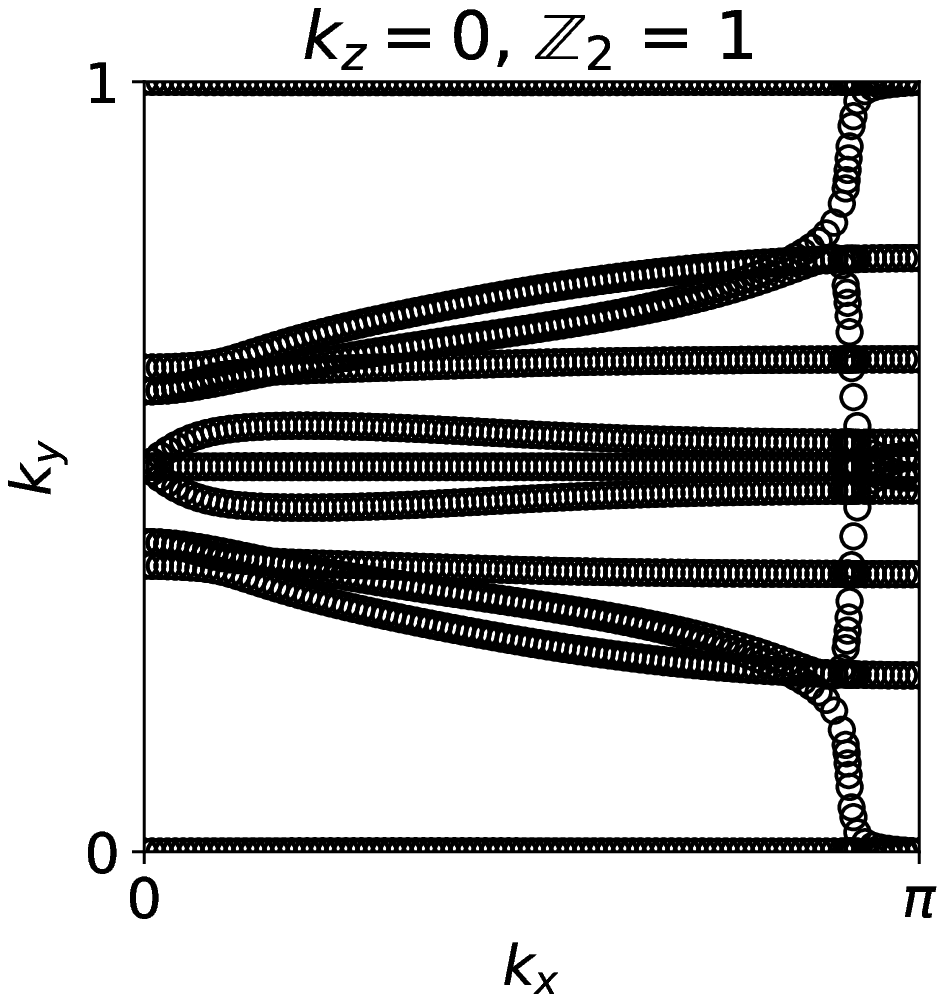}
        \end{subfigure}
        \hfill
        \begin{subfigure}[b]{0.24\textwidth}
                \caption{}
                \label{ycz2b}
                \includegraphics[width=\textwidth]{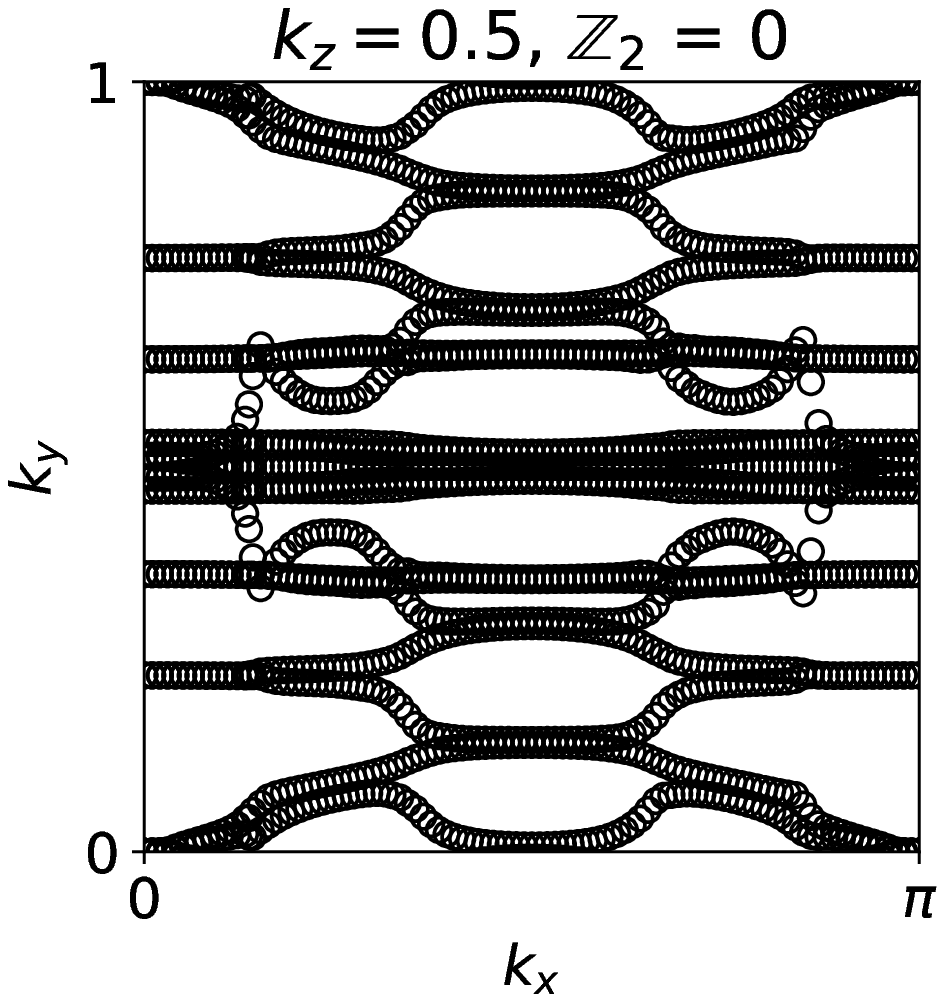}
        \end{subfigure}
        \hfill
        \begin{subfigure}[b]{0.24\textwidth}
                \caption{}
                \label{crcz2a}
                \includegraphics[width=\textwidth]{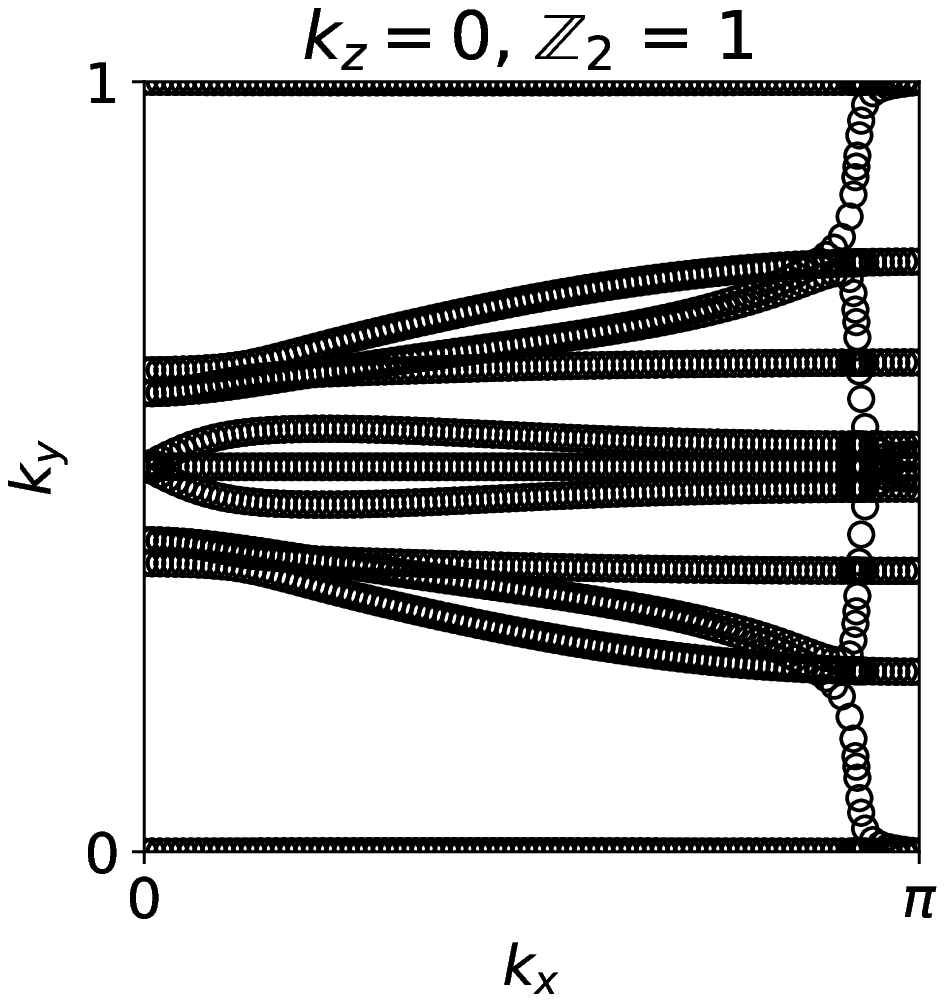}
        \end{subfigure}
        \hfill
        \begin{subfigure}[b]{0.24\textwidth}
                \caption{}
                \label{crcz2b}
                \includegraphics[width=\textwidth]{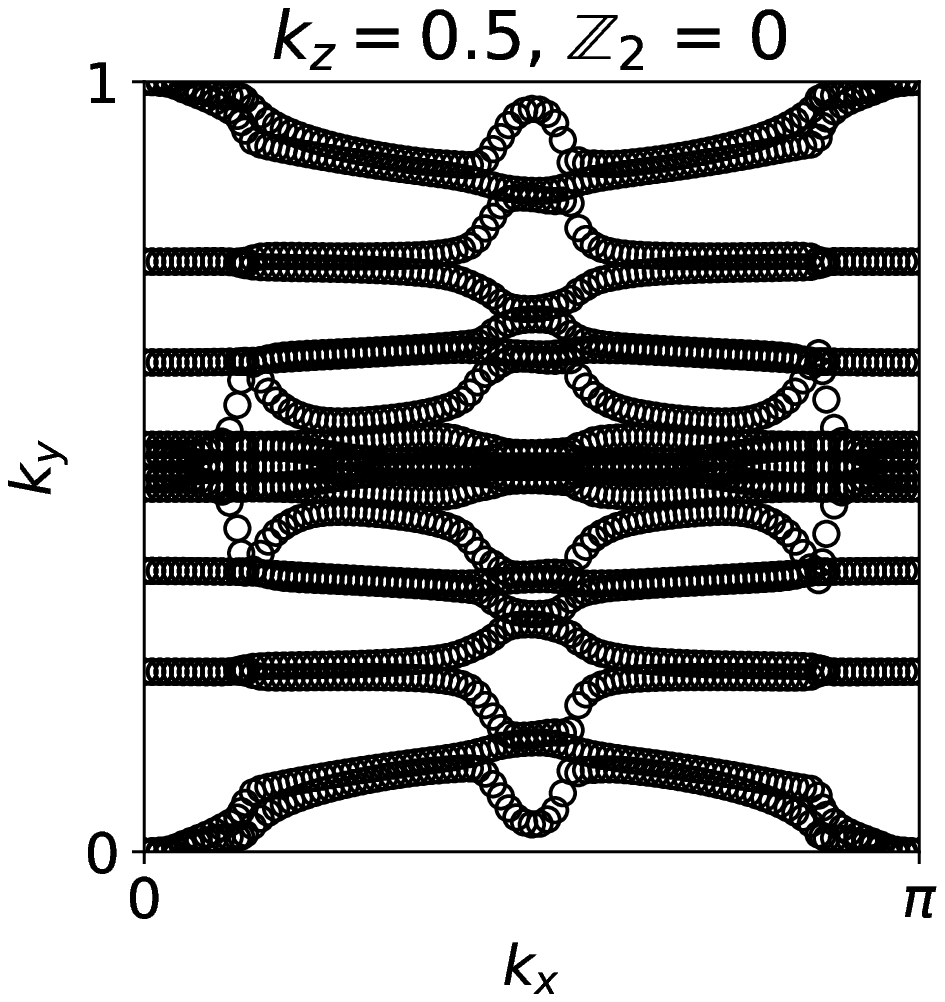}
        \end{subfigure}
        \\
        \begin{subfigure}[b]{0.24\textwidth}
                \caption{}
                \label{nbcz2a}
                \includegraphics[width=\textwidth]{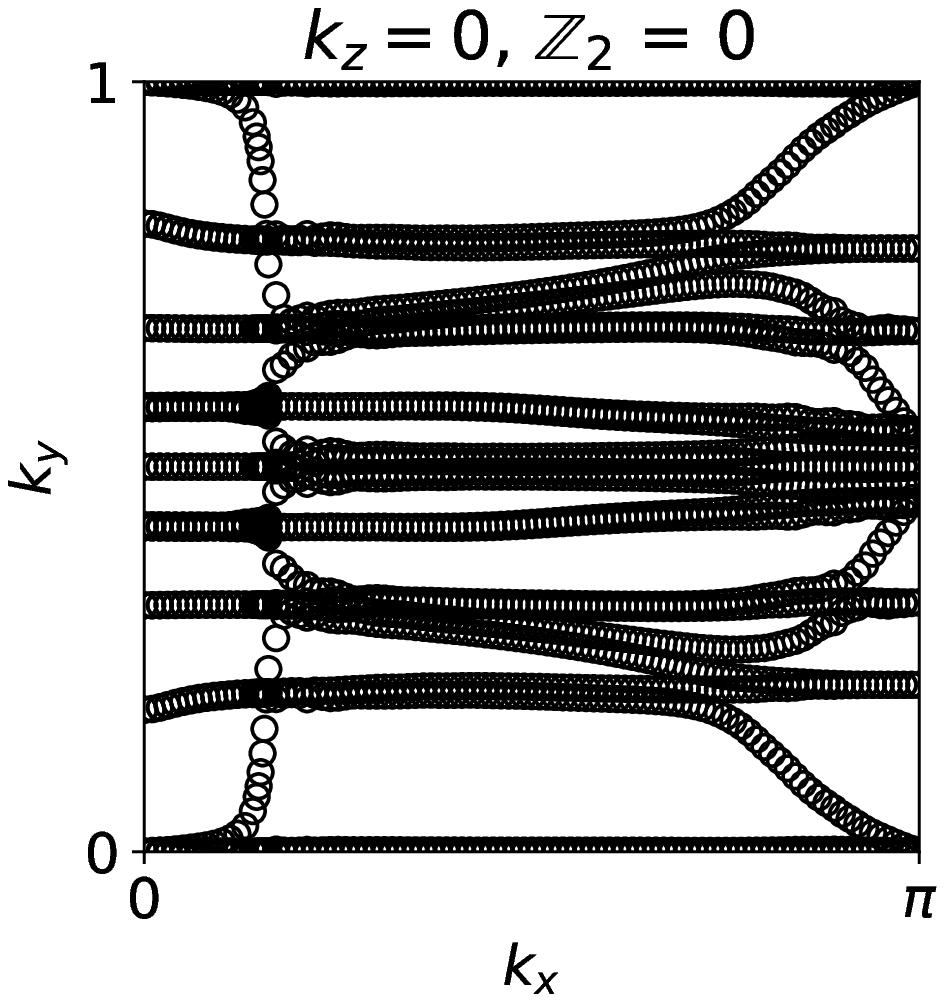}
        \end{subfigure}
        \begin{subfigure}[b]{0.24\textwidth}
                \caption{}
                \label{nbcz2b}
                \includegraphics[width=\textwidth]{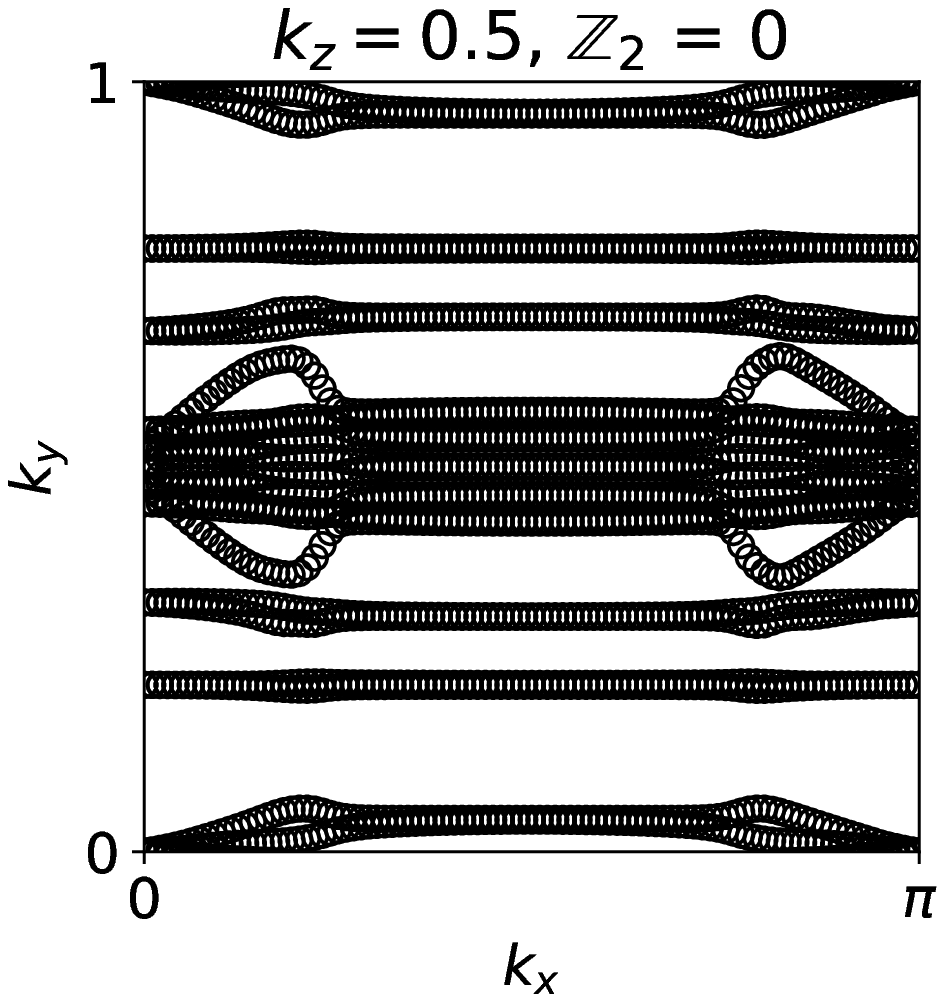}
        \end{subfigure}
        \begin{subfigure}[b]{0.24\textwidth}
                \caption{}
                \label{tacz2a}
                \includegraphics[width=\textwidth]{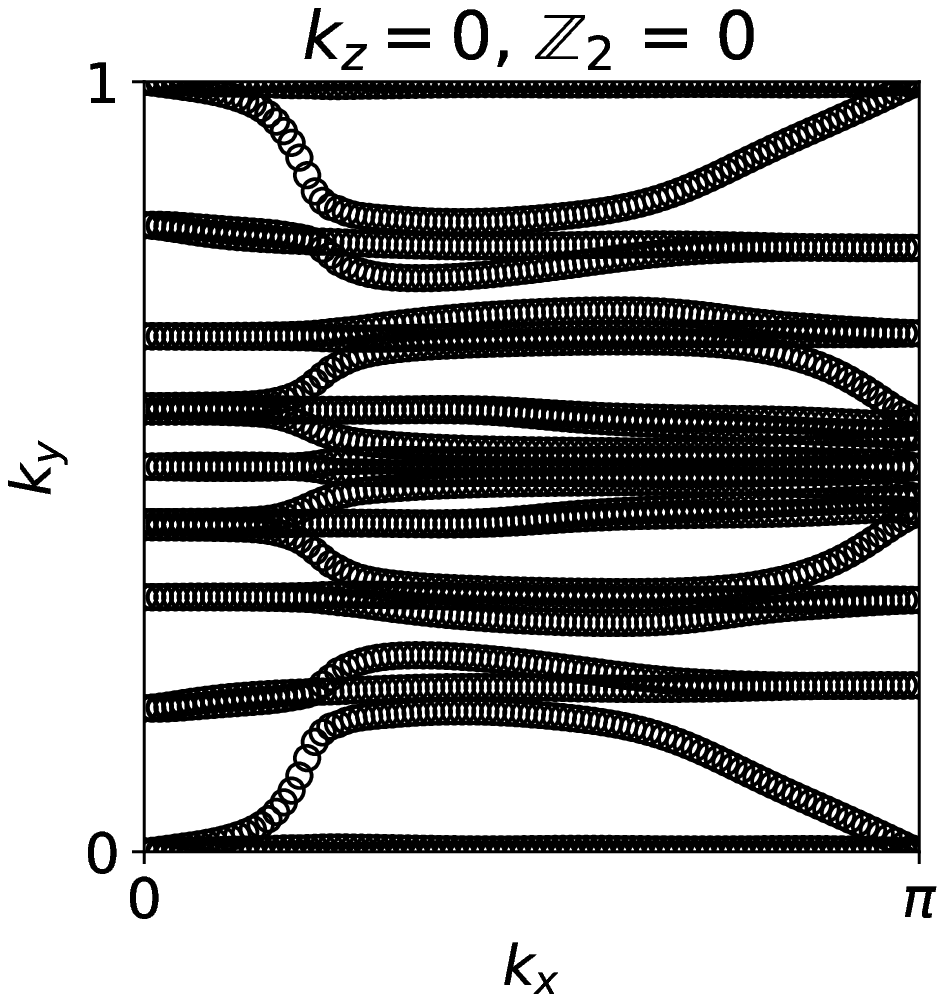}
        \end{subfigure}
        \begin{subfigure}[b]{0.24\textwidth}
                \caption{}
                \label{tacz2b}
                \includegraphics[width=\textwidth]{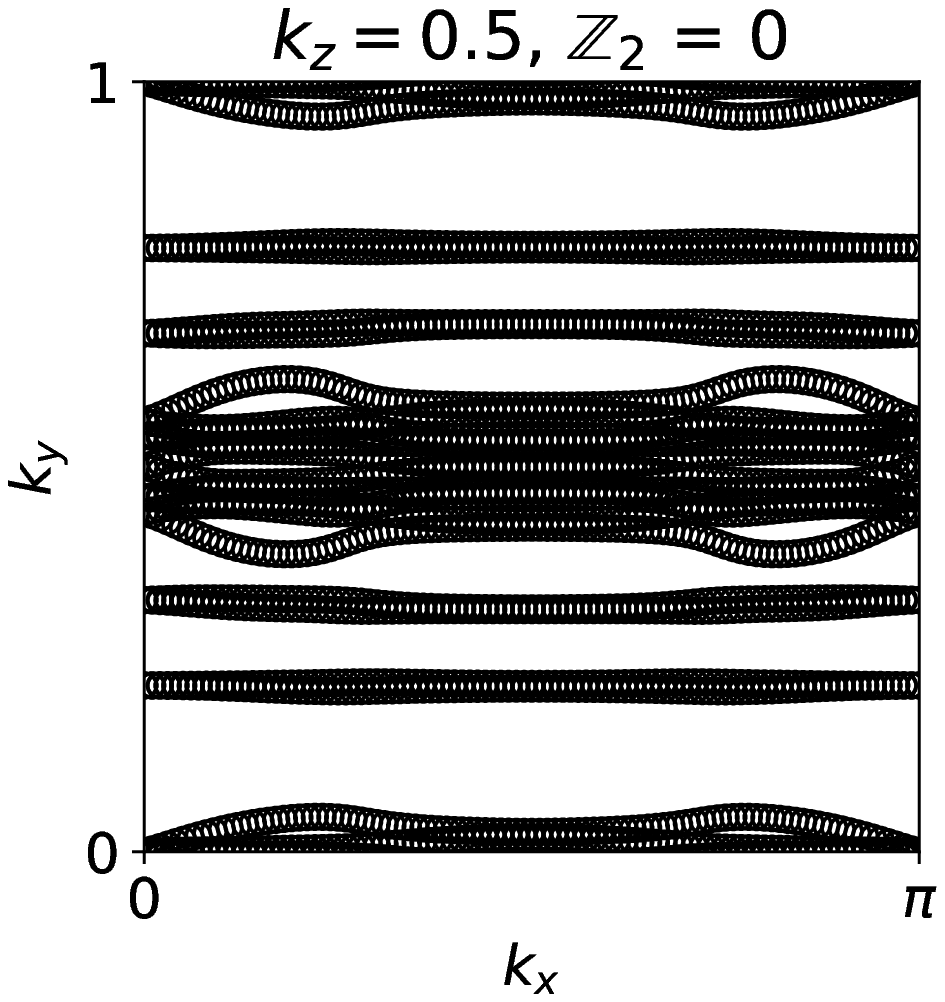}
        \end{subfigure}
        \caption{The black circles depict the Wannier charge centers.
        (a) The Wilson loop of VC at the surface $k_z = 0$ produces $Z_2 = 1$.
        (b) Wilson loop of VC at the surface $k_z = 0.5$ produces $Z_2 = 0$.
        (c) Wilson loop of CrC at the surface $k_z = 0$ produces $Z_2 = 1$.
        (d) Wilson loop of CrC at the surface $k_z = 0.5$ produces $Z_2 = 0$.
        (e) Wilson loop of NbC at the surface $k_z = 0$ produces $Z_2 = 0$.
        (f) Wilson loop of NbC at the surface $k_z = 0.5$ produces $Z_2 = 0$.
        (g) Wilson loop of TaC at the surface $k_z = 0$ produces $Z_2 = 0$.
        (h) Wilson loop of TaC at the surface $k_z = 0.5$ produces $Z_2 = 0$.}
        \label{z2}
\end{figure*}

\begin{figure*}[ht]
        \centering
        \begin{subfigure}[b]{0.49\textwidth}
                \caption{}
                \label{vcphonon}
                \includegraphics[width=\textwidth]{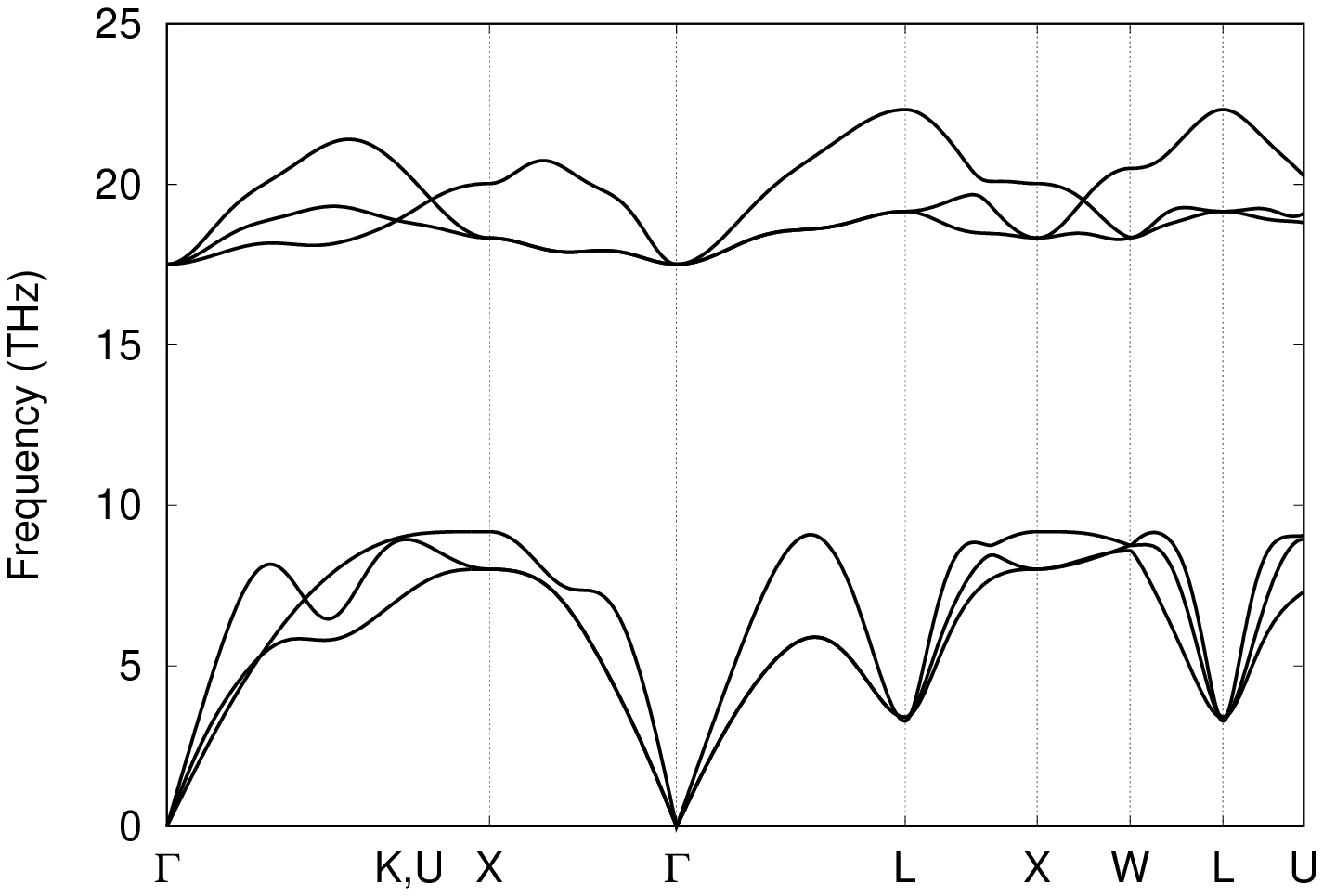}
        \end{subfigure}
        \hfill
        \begin{subfigure}[b]{0.49\textwidth}
                \caption{}
                \label{crcphonon}
                \includegraphics[width=\textwidth]{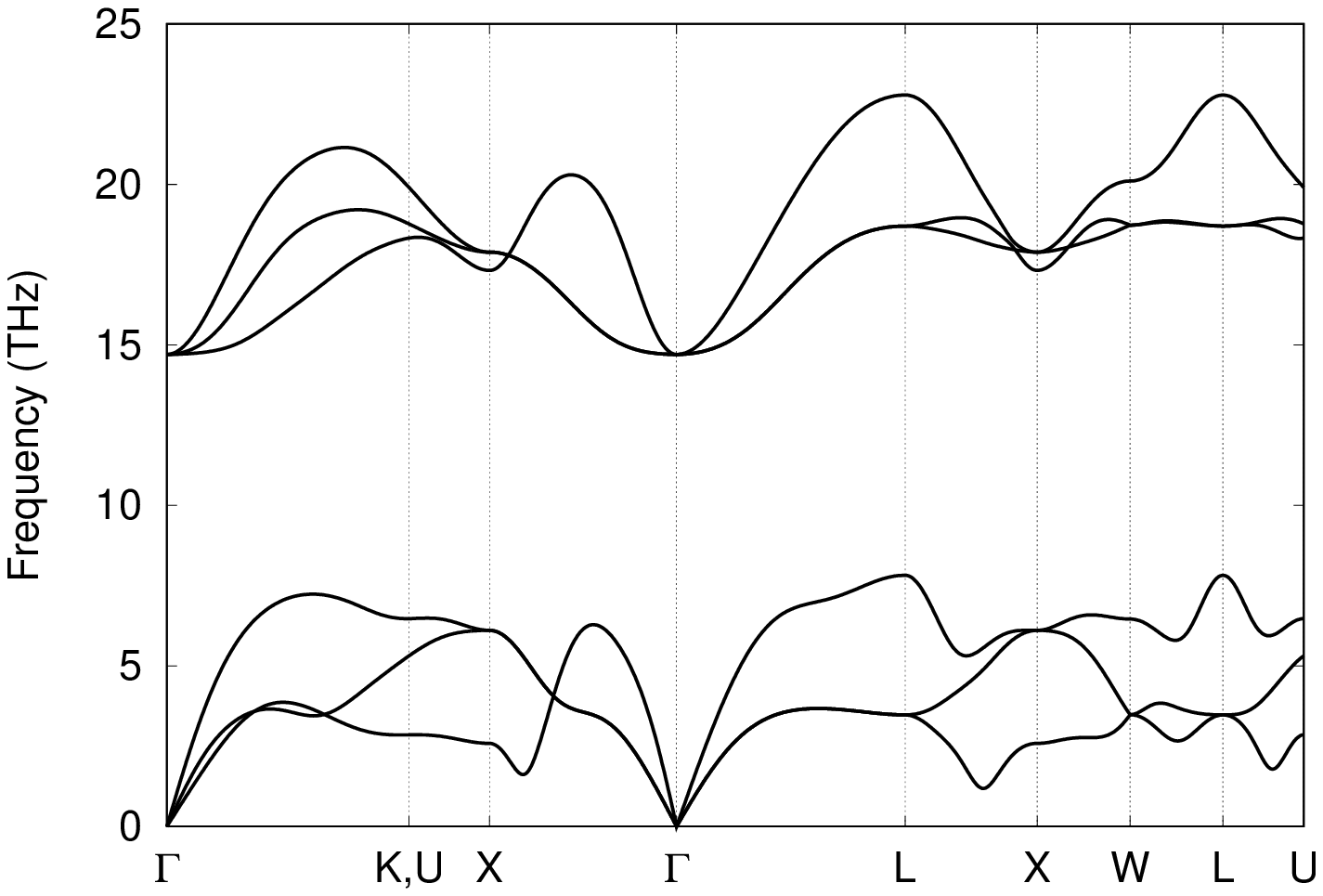}
        \end{subfigure}
        \caption{
                (a) Phonon dispersion of VC.
                (b) Phonon dispersion of CrC.
        }
\end{figure*}

\section{Results and Discussion}

We investigated the properties of various transition metal carbides including VC, CrC, TaC, and NbC.
First, we conducted total energy convergence tests on each of their energy cutoffs and k-point grids.
Then, we performed relaxation on their lattice parameters.
Afterwards, the band structures, densities of states (DOS), and partial densities of states (PDOS) with and without SOC were calculated.
We compared our band structures without SOC to other studies, since previous research generally did not include relativistic effects in their calculations.
The Wilson loops and $\mathbb{Z}_2$ topological invariants were calculated for each material as well.
We calculated the phonon dispersions of each compound to determine their stability.
Finally, we discussed recent advances in the synthesis of VC and CrC.

\subsection{Lattice Parameters}

We performed BFGS optimization on the lattice parameters of each material’s primitive cell.
Since the atoms in rock-salt structures are fixed due to high symmetry, we used a potential residual tolerance of $10^{-12}$ Ha as our stopping criteria for the self-consistent field cycles during relaxation.
Table \ref{relax} compares experimental determined lattice parameters with our theoretically calculated lattice parameters.
Our GGA parameters were in very good agreement with experimental values.
The lattice parameters calculated with LDA were underestimates as expected.
The reduced atomic coordinates of the rock-salt structure are (0, 0, 0) and (0.5, 0.5, 0.5).
Both atoms in the rock-salt structure can serve as inversion centers.

\begin{table}[ht]
        \centering
        \caption{Experimental vs. our theoretically calculated lattice parameters for each material.}
        \begin{ruledtabular}
                \begin{tabular}{L{0.25\mywidth}C{0.35\mywidth}C{0.2\mywidth}C{0.2\mywidth}}
                        Compound & Experimental ({\AA}) & GGA ({\AA}) & LDA ({\AA}) \\
                        \hline
                        VC & 4.16294\footnotemark[1] & 4.162 & 4.094 \\
                        CrC & 4.03\footnotemark[2] & 4.064 & 4.010 \\
                        NbC & 4.45443\footnotemark[1], 4.470\footnotemark[3] & 4.484 & 4.426 \\
                        TaC & 4.45300\footnotemark[1], 4.457\footnotemark[4] & 4.477 & 4.412 \\
                \end{tabular}
        \end{ruledtabular}
        \footnotetext[1]{Reference \citenum{nakamura2008crystal}.}
        \footnotetext[2]{Reference \citenum{liu1992a}.}
        \footnotetext[3]{Reference \citenum{storms1959the}.}
        \footnotetext[4]{Reference \citenum{bowman1961the}.}
        \label{relax}
\end{table}

\subsection{Electronic Band Structure}

Figure \ref{vcall} depicts the band structure of VC, not including the effects of SOC.
This band structure was very similar to previously reported results \cite{maibam2010electronic}.
Our DOS is also in very good agreement with earlier studies \cite{maibam2010electronic, zhukov1987the, neckel1975results}.
The DOS diagram shows that VC is metallic, like most transition metal carbides.
The regions near the Fermi level are almost completely dominated by the V-p orbitals, aside from a small contribution by the C-p orbitals between 2-2.5 eV.
The band calculation exhibits two band crossings in the $X-W$ region as shown by the red circles in Figure \ref{vcxw}.
These two nodal points signal the presence of a nodal ring, which is unstable when the effects of SOC are considered \cite{chiu2016classification}.
Including SOC in our band structure calculation opens up gaps between the the band crossings at $X-W$.
Most other regions of the band structure change only slightly after the inclusion of SOC since both vanadium and carbon atoms produce very weak SOC.
Since the nodal ring is gapped by SOC, it is very likely that this material is topologically nontrivial.
In later sections, we support this initial prediction by calculating the $\mathbb{Z}_2$ topological invariant.
In order to utilize the topologically nontrivial bands in experimental studies, VC would have to be doped approximately 3.09 electrons per primitive cell.

CrC's band structure without SOC is found in Figure \ref{crcall} and is extremely similar to that VC's band structure.
It compares well with other studies \cite{maibam2012investigation, tutuncu2012electrons}.
CrC's DOS reveals that it is also a metal, and its metallic properties arise from the transition metal atoms' d-orbitals.
When compared to VC, the additional d-orbital electron in Cr results in a downward shift of the DOS.
This shift leads to a higher DOS at the Fermi level in CrC.
Both our DOS and PDOS agree with earlier calculations \cite{maibam2012investigation, tutuncu2012electrons, singh1992electronic}.
Similar to VC, there are two band crossings between X and W that indicate the existence of a nodal ring (red circles in Figure \ref{crcxw}).
When the effects of SOC are included, the overall band structure remains similar to the band structure without SOC due to the weak SOC of the chromium and carbon atoms (Figure \ref{crcsocall}).
Even so, the nodal ring is gapped by SOC (Figure \ref{crcsocxw}).
Consequently, CrC may have nontrivial band topology.
Doping CrC with 2 electrons per primitive cell would raise the Fermi level, so the topologically nontrivial bands could be accessed.

The band structure of NbC without SOC is depicted in Figure \ref{nbcall}.
It is consistent with previous studies \cite{amriou2003fp-lapw, sharma2010studies}.
Since there is a large amount of overlap between the valence and conduction bands, NbC is clearly a metal.
Other studies' DOS are also comparable to ours \cite{amriou2003fp-lapw, schwarz1975partial, schwarz1977the}.
When SOC is not taken into account, NbC's $11^{\text{th}}$ and $12^{\text{th}}$ bands (shown by the red line) are degenerate along $\Gamma-X$ (Figure \ref{nbcgxw}).
Thus, the band structure of NbC without SOC does not have a band gap.
Once again, including the weak SOC of the niobium and carbon atoms in our band calculations changes little in the overall band structure (Figure \ref{nbcsocall}).
A small band gap opens in the previous band degeneracy at $\Gamma-X$ and $X-W$, leading to a continuous region with direct gaps (Figure \ref{nbcsocgxw}).
However, NbC's band structure with SOC does not show gapping along a nodal ring, so we cannot determine whether NbC is topologically nontrivial from the band structure. 
As a result, we use Z2Pack to determine whether NbC possesses a topologically nontrivial band structure.

Figure \ref{tacall} shows that TaC's band structure without SOC has many similarities to NbC.
This is consistent with earlier research \cite{sahnoun2005fp-lapw, sahnoun2005electronic}.
TaC has a region of suppressed DOS near 4 eV which also occurred in other studies \cite{sahnoun2005fp-lapw, sahnoun2005electronic}.
The $11^{\text{th}}$ and $12^{\text{th}}$ bands intersect in the same regions as NbC (shown by the red line in Figure \ref{tacgxw}).
Including SOC in our band structure calculation causes the degeneracy to become gapped (Figure \ref{tacsocgxw}).
A previous study also found the same effect in TaC \cite{sahnoun2005electronic}.
Since Ta atoms have stronger SOC than Nb atoms, the SOC-induced gapping in TaC is much greater than the gapping in NbC (Figure \ref{tacsocall}).
Similar to NbC, we cannot determine whether TaC is topologically nontrivial solely based on the band structure since it does not show gapping along a nodal ring.
Thus, we must calculate TaC's $\mathbb{Z}_2$ invariant to check if it is topologically nontrivial.

\subsection{Topological Invariance}

The Wilson loop method can also be used to calculate the $\mathbb{Z}_2$ topological invariant of both centrosymmetric and noncentrosymmetric structures with no direct band gap closures.
Figure \ref{z2} displays the Wilson loops of VC, CrC, NbC, and TaC in the planes $k_z = 0$ and $k_z = 0.5$.
These materials all contain a threefold rotational symmetry axis along the [111] direction.
Since the three weak invariants for each material are equivalent by symmetry, the Wilson loop only needs to be calculated on the $k_z = 0$ and $k_z = 0.5$ surfaces to calculate the strong index and three weak indices \cite{soluyanov2011wannier}.
Using Z2Pack, we calculated that in VC and CrC an odd number of Wannier charge centers (WCCs) are crossed when traversing the path created by the centers of the largest WCC gaps on $k_z = 0$ and an even number of WCCs are crossed for $k_z = 0.5$.
Thus, VC and CrC are likely both strong topological materials (1;000).
However, for NbC and TaC, their Wilson loops showed an even number of crossings for both $k_z = 0$ and $k_z = 0.5$.
Consequently, both of these materials are topologically trivial materials (0;000).

\subsection{Phonon Dispersion and Superconductivity}

Figure \ref{vcphonon} depicts the phonon band dispersion of VC.
Our VC phonon band diagram has no imaginary components, which implies that rock-salt VC is dynamically stable. 
This is consistent with experimental synthesis of VC \cite{schwartzkopf1953refractory, yazawa2004effect, ma2009low, chen2011a, hassanzadeh2015fast}.
In addition, the softening of the longitudal acoustic branches from $\Gamma-K$ confirms the superconductive nature of VC since peculiarities in the phonon spectrum are often associated with superconductivity.
Another study that calculated the phonon dispersion of VC found similar features, apart from the triple degeneracy at our L point compared to their double degeneracy \cite{isaev2007phonon}.
The aforementioned study also predicted that VC's T$_{\text{c}}$ was 11.5 K, a relatively high value compared to known topological superconductors. 

The phonon dispersion of CrC is shown in Figure \ref{crcphonon}.
The lack of imaginary frequencies suggests that rock-salt structure CrC is stable. 
Our calculated phonon dispersion has the same features as previous studies aside from a small discrepancy at the X point \cite{tutuncu2012electrons}.
Our quantitative values are also fairly accurate.
We calculated the transverse optical frequency at $\Gamma$ to be 14.702 THz, compared to 13.49 THz and 15.6 THz from other studies \cite{tutuncu2012electrons, singh1992electronic}.
Finally, CrC's soft acoustic branches likely cause it to have such a high predicted T$_{\text{c}}$ of 25-39 K \cite{kavitha2016structural, tutuncu2012electrons}. 
This extremely high T$_{\text{c}}$ greatly enhances the practicality of using CrC as a topological superconductor.

Advances in synthesizing both VC and CrC have been made in recent years.
In 2015, the mechanochemical combustion method was used to synthesize VC nanopowders \cite{hassanzadeh2015fast}.
Milling V$_2$O$_5$ with active carbon and magnesium for 80 minutes produced rock-salt VC without the need of additional heat treatment.
This approach was a significant improvement over the traditional high temperature methods of synthesizing VC since it was both faster and more economical \cite{schwartzkopf1953refractory, yazawa2004effect}.
Therefore, we believe this method may be a good choice to synthesize VC for future research in the field of topological superconductivity.
In regards to the synthesis of CrC, conventional high temperature reactions have struggled to produce metastable rock-salt chromium carbide \cite{liu1992a}. 
However, phase pure rock-salt CrC was recently realized through a low temperature salt flux synthetic method \cite{schmuecker2017synthesis}.
This advance grants future experimental studies access to this rare phase of CrC.

\section{Conclusion}
In summary, we have used first-principle calculations to theoretically predict the presence of topologocially nontrivial band structures in VC and CrC.
The SOC-induced gapping in each of these material's band structures suggests nontrivial band topology.
Additionally, the Wilson loops for each material were calculated.
Both VC and CrC had nonzero $\mathbb{Z}_2$ topological invariants while NbC and TaC had $\mathbb{Z}_2$ topological invariants of 0.
We also calculated the phonon band structures of VC and CrC and found that both materials were dynamically stable since they displayed no imaginary frequencies.
Moreover, the softening of acoustic modes present in the phonon dispersions confirms their superconductivity.
Therefore, VC and CrC display great promise as platforms for future research in the field of topological superconductivity.

\section*{Acknowledgements}
We thank Dr. Gefei Qian for providing technological support.
We also greatly appreciate Prof. Shengbai Zhang for helpful discussions.

\newpage
\bibliography{RichardZhanPaper}{}

\end{document}